\PassOptionsToPackage{dvipsnames}{xcolor}
\documentclass[twocolumn]{aastex631}
\usepackage{graphicx} % Required for inserting images
\usepackage{xcolor}
\usepackage{array}
\usepackage{multirow}
\usepackage{longtable}
\usepackage{appendix}

\begin{document}

\title{Hydrolyzed Hazes on Water-rich Exoplanets: Composition and Prebiotic Chemistry}

\author[0009-0007-6910-6347]{Cara Pesciotta}
\affil{Department of Earth and Planetary Sciences, Johns Hopkins University, Baltimore, MD, USA}

\author[0000-0003-4596-0702]{Sarah M. H{\"o}rst}
\affil{Department of Earth and Planetary Sciences, Johns Hopkins University, Baltimore, MD, USA}

\author[0000-0002-4072-181X]{Michael J. Radke}
\affil{Department of Earth and Planetary Sciences, Johns Hopkins University, Baltimore, MD, USA}

\author[0000-0002-6721-3284]{Sarah E. Moran}
\affil{Department of Astronomy, University of Maryland, College Park, MD, USA}

\author[0009-0005-4064-5916]{B.P. Blakley}
\affil{Department of Earth and Planetary Sciences, Johns Hopkins University, Baltimore, MD, USA}

\author[0000-0002-6694-0965]{Chao He}
\affil{School of Earth and Space Sciences, University of Science and Technology of China, Hefei, China}

\author[0000-0001-7273-1898]{V{\'e}ronique Vuitton}
\affil{Univ. Grenoble Alpes, CNRS, IPAG, 38000 Grenoble, France}

\author{Laur{\`e}ne Flandinet}
\affil{Univ. Grenoble Alpes, CNRS, IPAG, 38000 Grenoble, France}

\author{Kathy Dubuc}
\affil{Univ. Grenoble Alpes, CNRS, IPAG, 38000 Grenoble, France}

\begin{abstract}
Although photochemical hazes complicate the spectroscopic analysis of sub-Neptunes, the very presence of hazes can teach us about the habitability of these planets. Aside from affecting a planet's atmospheric and surface dynamics, hazes offer a source of organics potentially including prebiotic molecules like amino acids, nucleobases, and sugars. These organics are particularly compelling for temperate, water-rich planets where hazes may react with liquid water in condensate clouds or surface reservoirs. Hydrolysis reactions drive molecular reorganization and oxygenation, increasing the chemical complexity of the prebiotic inventory. This water-based evolution of hazes in their environment is largely unexplored and will have consequences for their chemical and optical behavior. In this work, we examine the chemical evolution of temperate water-rich exoplanet haze analogs in liquid water. High-resolution mass spectrometry finds thousands of CHNO molecules with bonding behaviors and chemical complexities distinct from the unaltered haze analogs. The oxygen weight percent increases significantly in the soluble fraction and moderately in the insoluble fraction after hydrolysis, with different carbon bonding structures demonstrating the redistribution of functional groups. The hydrolyzed samples exhibit a broader distribution in H/C and O/C ratios, and the molecular formulas suggest complex carbohydrate chemistry that challenges current theories of prebiotic organic delivery. 
\vspace{1cm}
\end{abstract}

\section{Introduction}
A key challenge in astrobiology, particularly for exoplanets, is the ability to link observations to signs of life and habitability. Many studies focus on what makes a habitable environment (e.g. liquid water, temperate surfaces, low XUV flux), while others are concerned with biosignature detection (false positives, false negatives, and the ambiguities between them). Spectroscopy from ground- and space-based observatories seeks to reveal atmospheric molecular features indicative of these phenomena. Water, for example, has been tentatively observed in the atmospheres of several temperate sub-Neptune exoplanets, defined as planets 1.75--4 R$_E$ with temperatures $<$1000 K. Some models predict global water oceans, as for planets like K2-18b \citep{Benneke2019,Madhusudhan2023}, or steamy atmospheres, as on GJ 9827d \citep{Piaulet2024} and TOI-270d \citep{Benneke2024}. Many sub-Neptunes, however, have muted or even flat transmission spectra that complicate analysis (e.g., \citealt{Kreidberg2014,Knutson2014a,Dragomir2015,Kempton2023,Gao2023,Ohno2025}). This lack of identifiable spectral features precludes understanding their composition, formation, and habitability.

 Muted transmission features can be signs of high-altitude aerosols in the atmosphere that can both obscure gas-phase species from detection and influence planetary habitability. The most common planetary aerosols are clouds and hazes. Their presence alters the radiation budget, atmospheric chemistry, and material transport, having a demonstrable effect on a planet's habitability (e.g., \citealt{Marley2013,Lavvas2021,Steinrueck2025}). Aerosols regulate temperature by cooling through the scattering of starlight or warming through the absorption of stellar and thermal irradiation, potentially expanding the habitable zone \citep{Toon1980,Gao2021}. UV-absorptive aerosols may also shield the planet from intense irradiation, protecting the atmosphere and surface from destructive stellar flares \citep{Tilley2019}. Furthermore, hazes themselves are reservoirs of prebiotic compounds. Formed through photolysis of atmospheric gases, hazes consist of complex organic molecules assembled from atmospheric C, H, N, O, or other elements. Laboratory synthesis of analog haze material suggests amino acids, nucleobases, and potentially even sugars in their composition (e.g., \citealt{Khare1984,Horst2012,Kawai2019,Moran2020,Vuitton2021,Moran2022,Pearce2024a,Reed2025}).

The formation of prebiotic molecules in atmospheres presents an additional avenue for preliminary abiogenesis alongside external delivery or hydrothermal systems. Observations and laboratory analog experiments of comets, asteroids, and interstellar ices reveal prebiotic molecules (e.g., \citealt{MunozCaro2013,Kaiser2015,Altwegg2016,Arumainayagam2019,Sandford2020,Mojarro2025}), while models consider whether the molecules can survive atmospheric entry (e.g., \citealt{Pierazzo1999,Todd2020,Anslow2023}). Theoretical, laboratory, and field investigations into hydrothermal systems also demonstrate the formation and stability of these types of molecules, leveraging fluid-rock interactions as natural catalysts and geothermal heat as an energy source \citep{Martin2008,Mielke2010,Sojo2016,Barge2022}. Current theories rely on these methods for creating and preserving building block materials. The presence of prebiotic molecules in hazes, however, suggests that atmospheres alone can create the building blocks for life \citep{Pearce2024a,Reed2025}. Determining how efficiently these molecules survive and accumulate after formation will be critical for assessing the contribution of atmospheric chemistry to prebiotic chemistry.

As sources of prebiotic material, it is important to understand how hazes evolve in their environment after formation. At high altitudes, hazes may be subject to stellar irradiation that alters the chemical structure \citep{Carrasco2018,Gavilan2018,Huseby2025} or potentially degrades prebiotic molecules (e.g., \citealt{tenKate2005,Zetterlind2025}). Gravitational settling may protect the molecules from these effects. The physical properties of laboratory-produced haze analogs also favor efficient cloud condensation nucleation and surface settling \citep{Yu2021}, allowing haze particles to exist in many types of planetary environments. After formation, transport to lower atmospheric layers may bring hazes to altitudes with liquid water clouds \citep{Kawashima2018,Lavvas2026}, where they can act as cloud condensation nuclei (CCN; \citealt{Mang2022,Lavvas2026}) and later rainout or regenerate should the cloud evaporate \citep{Ervens2015,Arieli2025}. Continued transport may eventually deliver a fraction of the particles to surface reservoirs. In particular, these water-rich conditions such as the interfaces of clouds, raindrops, lakes, and seas may have direct implications for early life \citep{Miller1957}. 

Reactions between haze and water may promote the breakdown and reassembly of organics into increasingly complex prebiotic species \citep{Ross2016}. Hydrolysis and condensation reactions are often considered essential for prebiotic chemistry because they drive molecular diversity \citep{Miller1957,Mamajanov2014,Becker2018}, and previous laboratory work relevant for Saturn's moon Titan demonstrates that hydrolysis can provide the oxygen required for the synthesis of prebiotic molecules \citep{Khare1986,Neish2010,Poch2012,Cleaves2014}. Aqueous aerosol interfaces are shown to act as prebiotic reactors, increasing biomolecule detection under certain conditions \citep{Mompean2019}.

In the quest to observe prebiotic processes like these on exoplanets, spectra consistent with flat lines are often seen as a hindrance. We cannot constrain gaseous abundances, infer surface conditions, or even determine if an atmosphere is present on the planet. In relation to photochemistry, however, flat lines do not rule out a hazy world full of organic chemistry. \cite{Pesciotta2026} finds that hydrolyzed hazes are significantly more absorbing than their unaltered counterparts and result in flatter transmission spectra than equivalent particle loading of fresh hazes, demonstrating that rich chemistry may be hiding under colorless spectroscopy. This work presents the chemistry behind laboratory hydrolysis experiments on exoplanet haze analogs as a complementary study to the optical properties presented in \cite{Pesciotta2026}. We use high-resolution mass spectrometry to determine the bulk composition, including elemental ratios, as well as exact masses of molecules created and destroyed after hydrolysis. Through comparison pre- and post-hydrolysis, we discuss the implications for prebiotic chemistry on water-rich exoplanets.

% biosignatures, from molecules solely produced by biology to how biology shapes an atmosphere, while others are concerned with their detection, from false positives to false negatives and things in between.

% Despite being the most abundant class of planet, the nature of sub-Neptune exoplanets is poorly understood. \textcolor{red}{Here's what we do know about them.} Outstanding questions include their composition, formation, and potential for habitability. Clues from bulk density imply either a silicate-iron core with a hydrogen-rich atmosphere or an icy core with a water-rich atmosphere, and planet formation theories suggest atmospheric escape or core-powered mass loss as the main drivers behind the two respective chemistries. Spectroscopy from ground- and space-based observatories has sought to reveal atmospheric composition, however muted or even flat transmission spectra complicate their analysis. Muted transmission features are signs of high-altitude aerosols in the atmosphere, obscuring gas-phase species and influencing the planetary environment. 

\section{Methods}
\label{sec:methods}
\subsection{Haze Analog Production \& Hydrolysis}
\begin{figure*}
    \centering
    \includegraphics[width=0.95\linewidth]{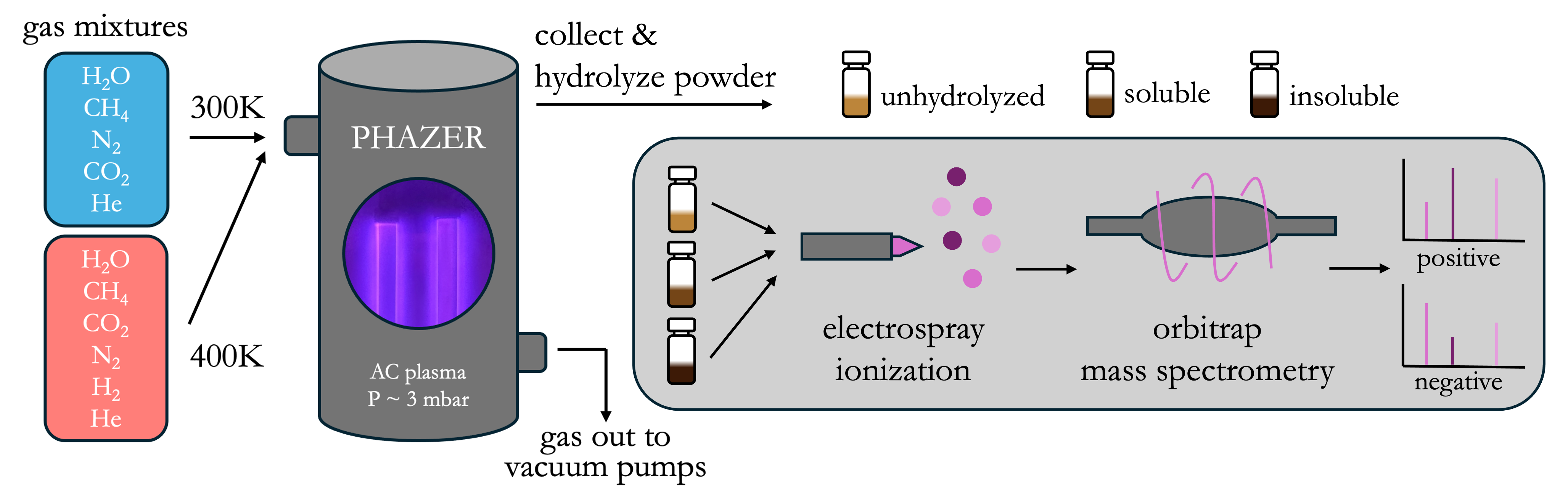}
    \caption{Simplified schematic of the experimental setup, hydrolysis procedure, and ESI Orbitrap. PHAZER haze analogs for the two simulated atmospheres react with water, producing soluble and insoluble fractions of the samples. We subject the three samples (unhydrolyzed, soluble, and insoluble) to electrospray ionization (ESI) and Orbitrap mass spectrometry in both positive and negative mode.}
    \label{fig:schematic}
\end{figure*}

Haze analogs are produced in the PHAZER chamber at Johns Hopkins University \citep{He2017}, designed to simulate photochemistry in a wide range of planetary atmospheres. The water-rich haze analogs analyzed here have been studied in detail in our previous investigations \citep{Horst2018,He2018,Moran2020,Yu2021,He2024,Pesciotta2026}, allowing our results to be placed in context. In-depth explanations of the experimental setup and procedures can be found therein, with relevant details included below. A schematic of the setup and procedure is shown in Figure \ref{fig:schematic}.

Initial gas mixing ratios are calculated from equilibrium chemistry models of 1000x solar metallicity atmospheres at 300 and 400 K (see Table \ref{tab:mixingratio}; \citealt{Moses2013}). We focus on these two cases because they combine high atmospheric H$_2$O abundances with high haze production rates, making the simulated atmospheres particularly favorable for aqueous processing of hazes. These premixed gases run through a heating coil to achieve the desired temperature and then flow through a stainless steel reaction chamber at a pressure of $\sim$3 mbar, simulating the conditions in upper atmospheres where photochemistry is most active. The gas is exposed to an AC glow discharge, producing charged particles analogous to cosmic ray and energetic electron bombardment but with energies from 1.5 to 18.5 eV that represent UV irradiation under an experimentally-feasible timeline \citep{Pearce2022}. Reactions form gas and solid-phase products inside of the chamber that vary in chemical and physical properties depending on the experimental conditions. The solid products, our haze analogs, are collected and stored inside of a dry N$_2$ glovebox to prevent contamination and reactions with ambient air.

\begin{table}[]
\centering
\caption{Initial gas mixtures, production rates, and upper limit solubilities for water-rich haze analogs. Values in parentheses indicate the corresponding upper limit percentage of soluble material.}
\begin{tabular}{l c c c}
    \hline \hline
    \multirow{2}{*}{Planet} & \multirow{2}{*}{Mixing Ratio} & Production & Solubility \\
     & & Rate (mg h$^{-1}$) & (mg/mL) \\
    \hline
    300 K & 66.0\% H$_2$O & 10.43 & 13.4 \\
    Water-rich & 6.6\% CH$_4$ & & (81.7\%) \\
    Exoplanet & 6.5\% N$_2$ & & \\
     & 4.9\% CO$_2$ & & \\
     & 16.0\% He & & \\
    \hline
    400 K & 56.0\% H$_2$O & 10.00 & 13.8 \\
    Water-rich & 11.0\% CH$_4$ & & (84.9\%) \\
    Exoplanet & 10.0\% CO$_2$ & & \\
     & 6.4\% N$_2$ & & \\
     & 1.9\% H$_2$ & & \\
     & 14.7\% He & & \\
\end{tabular}
\label{tab:mixingratio}
\end{table}

To study reactions with water, we adapt hydrolysis procedures from previous Titan haze hydrolysis experiments \citep{Neish2008,Ramirez2010,Poch2012,Cleaves2014}. We mix 30 mg of each haze analog sample with 2 mL HPLC grade water in a sealed glass vial. The solution reacts for three weeks at room temperature (294 K), and we periodically shake the vials to increase the surface area of sample reacting with water. Based on the prior experiments, these experimental parameters should allow for most chemical reactions to occur in the solution, however a time-resolved hydrolysis study would be required to properly quantify the reactions and their rates. After hydrolysis, the solution is centrifuged and split into soluble and insoluble fractions by pouring off the liquid into a separate vial. The fractions are stored in a fume hood as water, in addition to highly-volatile compounds in the samples, evaporates away until fully dry.

\cite{Pesciotta2026} presents the physical properties of the hydrolyzed haze analogs; see this companion paper for thorough discussions of the results, implications for cloud cover and observability, and sources of error. Throughout the experiment, the solutions change color from shades of orange to brown (see images in Fig. 1 of \citealt{Pesciotta2026}). After separation and drying, the soluble fractions are sticky and viscous while the insoluble fractions are solid and appear powdery. Solubility is calculated by subtracting the mass of collected sample from the initial sample mass and dividing by the volume of water used for hydrolysis, yielding an upper limit estimate due to limitations in sample recovery and potential volatile loss. We report solubilities of 13.4 and 13.8 mg/mL for the 300 and 400 K planets, respectively, indicating substantial aqueous processing and hydrolyzed hazes should be considered in theory and models. We note that we do not collect enough mass of the 400 K planet soluble fraction to measure the optical properties in our prior analysis. Importantly, the mass spectrometry technique employed in this study requires much less sample than the optical measurement, allowing us to measure the 400 K soluble fraction for the first time.

\subsection{Orbitrap Mass Spectrometry Measurements}
We perform high-resolution mass spectrometry with a Thermo Fisher Scientific LTQ-Orbitrap XL mass spectrometer with electrospray ionization (IPAG, Grenoble, France). The instrument achieves a resolving power better than $10^5$ between 200 and 400 m/z with exact mass accuracy of $\pm$2 ppm. We use an Orbitrap as opposed to other mass spectrometers because of the large spectral range and its ability to analyze complex mixtures. Electrospray ionization, a soft ionization method, allows for analysis of large molecules without fragmenting them. The methods are also compatible with small amounts of sample ($\sim$tens of $\mu$g), enabling measurement of low-yield hydrolysis products.

To prepare for measurement, samples must be dissolved in a polar solvent. While water itself is a common electrospray solvent, this would inhibit measurement of the insoluble fraction. Instead, we use methanol as the solvent for all samples \citep{Moran2020}. 1.5 mL of methanol is poured directly into the sample tubes ($<$1 mg sample). The mixture is sonicated for 1 hour and centrifuged for 5 minutes at 10,000 rpm. The solution is diluted again with methanol at a 1:1 ratio. Even using methanol rather than water, the insoluble fraction does not dissolve completely. As such, the mass spectrometry measurements represent the methanol-extractable molecular inventory rather than the complete sample. This may introduce compositional bias into the results, however identical extraction procedures ensure a consistent basis for comparison. Blanks are also prepared with pure methanol in a sample tube, measured before each sample measurement.

Measurements are taken across three mass bins (50–300 m/z, 150–450 m/z, and 400–1000 m/z) where overlap in the bins enables comparison across them. The tube lens voltage is set to 50, 60, and 90 V to optimize signal in each respective mass bin, and each sample scan is an average of 1024 microscans per mass bin. We measure mass spectra in both positive and negative ion modes to capture the different types of compounds observed in both modes. The spray voltage is set to 3.4 and 3.8 kV for positive and negative modes, respectively.

\subsection{Data Analysis}
The mass spectrometry data are extremely complex, containing thousands of peaks whose distribution and intensity are dependent on molecular functional groups and ionization efficiency. This warrants rigorous background subtraction and careful consideration of contaminants that may be in the data. We compare peaks present in the sample spectrum to each corresponding blank using a custom background subtraction algorithm. This algorithm takes into account the fact that intensities across Orbitrap measurements are not comparable, as the instrument aims to maximize the number of accumulated ions in each measurement. Blank peaks above an intensity threshold are removed from the sample spectra. We then plot the mass defect for the remaining peaks, calculated by subtracting a peak's nominal mass from the exact mass. The mass defect is a visual representation of repeating chemical families in the spectra, and it allows us to remove obvious instrument artifacts and contaminants (see Appendix \ref{sec:md} for more detail). Full mass spectrometry analysis (described below) for both the original spectra and cleaned spectra demonstrates that the cleaning procedure improves compatibility with the chain growth constraints incorporated into the analysis code.

Mass spectra are analyzed using the custom IDL/\- Fortran routine \textit{idmol} written for Titan haze analog analysis \citep{Horst2011} and modified to reflect exoplanet chemistry by expanding the allowed oxygen content of assigned molecular formulas \citep{Moran2020}. The code assigns molecular formulas to peaks in the data, taking into account user-specified parameters in addition to known molecular bonding patterns. We run \textit{idmol} separately for each measurement mass bin and instrument mode. It first calculates all possible CHNO molecules for a given mass and narrows down the options given a specified instrument mode, maximum number of oxygen atoms, mass tolerance, intensity threshold, and nitrogen-to-carbon ratio. The program then uses the nitrogen rule to further constrain assignments, verifying that molecules with an even nominal mass have an even number of nitrogen atoms and odd masses have an odd number of nitrogen atoms. Low mass assignments inform high mass assignments based on chain growth mechanisms, as non-unique formulas are possible greater than 300 m/z \citep{Gautier2014}. The mass spectrometry measurements for the original haze analogs from \citet{Moran2020} are reanalyzed using the post-processing workflow described here to enable direct comparison with the hydrolyzed samples.

We compare the resulting molecular assignments to the molecular formulas for common prebiotic molecules including amino acids, nucleobases, and sugars. We note that molecular structure cannot be determined by \textit{idmol} or with Orbitrap mass spectrometry alone. Another technique such as tandem mass spectrometry or NMR spectroscopy would be required to analyze the spatial arrangement of atoms. As such, comparisons to prebiotic molecules serve as a first order look at prebiotic chemistry in these samples.

We also perform elemental analysis of the identified molecular formulas. Elemental analysis is performed separately for the three measurement mass bins but combines positive and negative mode measurements. This captures a more complete view of the samples while leveraging the standard deviation between mass bins to estimate uncertainty in the elemental ratios. The calculated elemental ratios are intensity-weighted and use average atomic masses. Previous studies that employed this method in tandem with combustion analysis demonstrated that Orbitrap measurements tend to underestimate the intensity of oxygen-containing molecules, and the lowest 10\% of intensities should be boosted by a factor of 10 to be in agreement with results from combustion analysis \citep{Horst2011,Moran2020,Moran2022}. 

\section{Results}

\subsection{Mass Spectrometry}
\label{sec:ms}
\begin{figure*}[]
    \centering
    \includegraphics[width=0.95\textwidth]{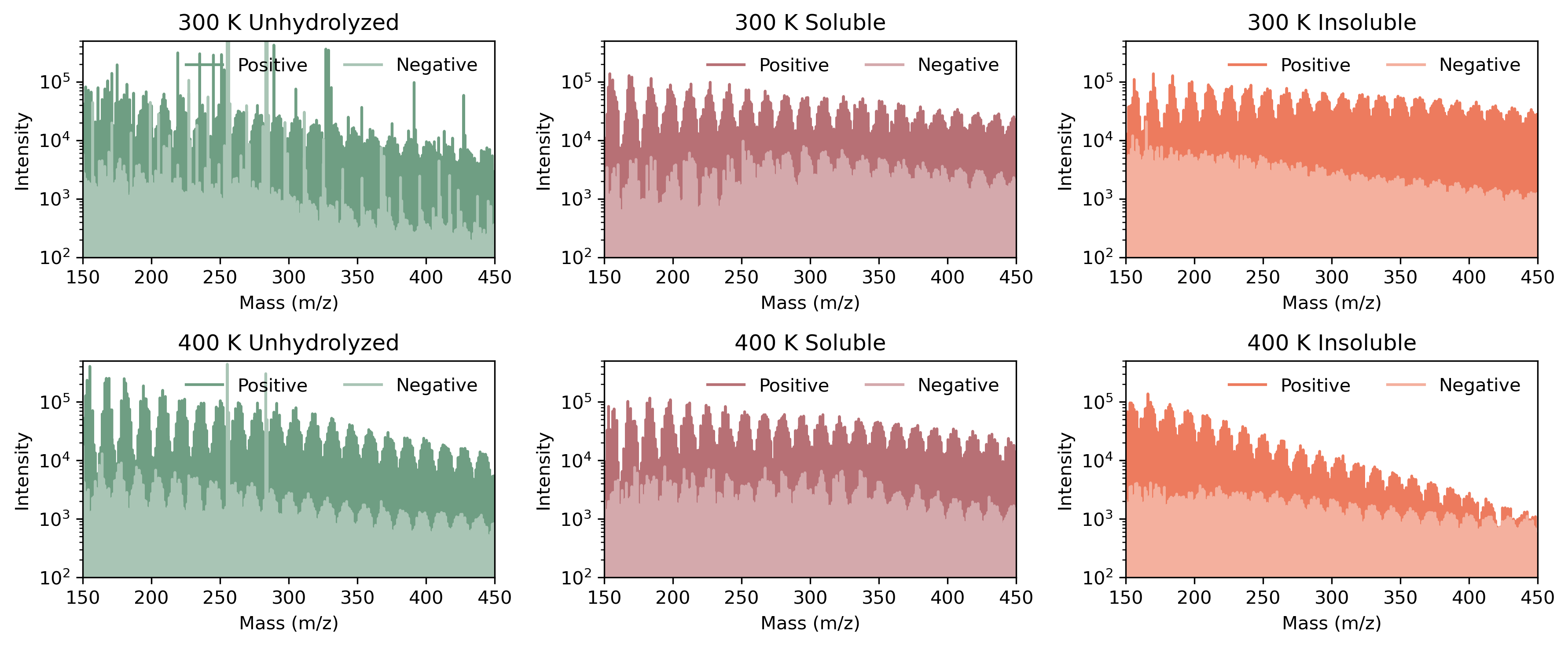}
    \caption{Mass spectra from 150--450 m/z of unaltered (left) and hydrolyzed (middle: soluble; right: insoluble) haze analogs for two water-rich exoplanets (top: 300 K; bottom: 400 K). For each sample, the positive mode measurement is plotted in a darker hue while the negative mode is plotted in a lighter hue.}
    \label{fig:spectra}
\end{figure*}

Figure \ref{fig:spectra} shows mass spectra from 150 to 450 m/z of hydrolyzed and unhydrolyzed haze analogs for the two water-rich exoplanets. All three mass bins are included in the analyses; however, only the 150-450 m/z bin is shown in Figure \ref{fig:spectra} as a representative range illustrating the overall spectral shape. Spectra for the low and high mass bins can be found in Appendix \ref{sec:addms}. The repeating bell-shaped envelopes across the spectra are common for organic samples, originating from repeating chemical families (e.g., \citealt{Gautier2014,Zark2018}). Spacing between peaks gives information about homologous series in the samples. All average peak spacings are between 13 and 14 amu, indicating that hydrolysis does not significantly affect the primary building block of the organic chains. While a spacing of 14 amu is commonly attributed to CH$_2$, adducts with various CHNO combinations can affect peak spacing with precision better than 1 amu (e.g., Titan haze's 13.5 amu spacing;  \citealt{Sarker2003,Pernot2010,Horst2011,Gautier2014}). As such, it is likely that the samples have different homologous series made from various combinations of CHNO. 

% (e.g., \citealt{Marshall2004})

The original 300 K sample has a distinct, sharp shape compared to the 400 K sample and the other haze analogs in \cite{Moran2020}. This may be due to differences in the initial gas mixture for haze production, perhaps the lack of H$_2$, leading to unique chemistry in the haze analogs. The mass defect of this sample demonstrates that the sharp peaks plot along a different homologous series than the typical bell-shaped envelopes (see Appendix \ref{sec:md}), and detailed analysis of the mass spectra peak spacing may distinguish the two chemical backbones. The disappearance of the sharp peaks after hydrolysis indicates the breakdown of those chemical families. Section \ref{sec:bulk} further speculates on this transformation using bulk chemical properties.

The intensity of a peak in the mass spectra does not solely correspond to the abundance of the molecule, but also reflects how well the molecule is ionized. Ionization efficiency depends on molecular functional groups, which influence whether the molecule ionizes in positive or negative mode and the intensity of the resulting signal. The positive mode measurement of the 400 K analog insoluble fraction displays a steep monotonic drop in intensity with increasing molecular weight that is not observed in the other samples. This might imply the sample is made of non-polar compounds, as they tend to resist protonation. This is consistent with surface energy measurements from \cite{Yu2021} finding that the 400 K analog is considerably less polar than the 300 K analog. More detail regarding the composition can be found in Section \ref{sec:bulk}.

\subsection{Bulk Chemistry}
\label{sec:bulk}
The molecular formulas assigned by \textit{idmol} are analyzed in bulk to identify population-scale trends in bonding behavior. These analyses can illuminate potential chemical pathways, including oxygen incorporation and polymerization, as well as link chemistry to observations using the optical properties measured in \cite{Pesciotta2026}.

\subsubsection{Elemental and Structural Analysis}
\label{sec:ea}
\begin{figure}[]
    \centering
    \includegraphics[width=0.4\textwidth]{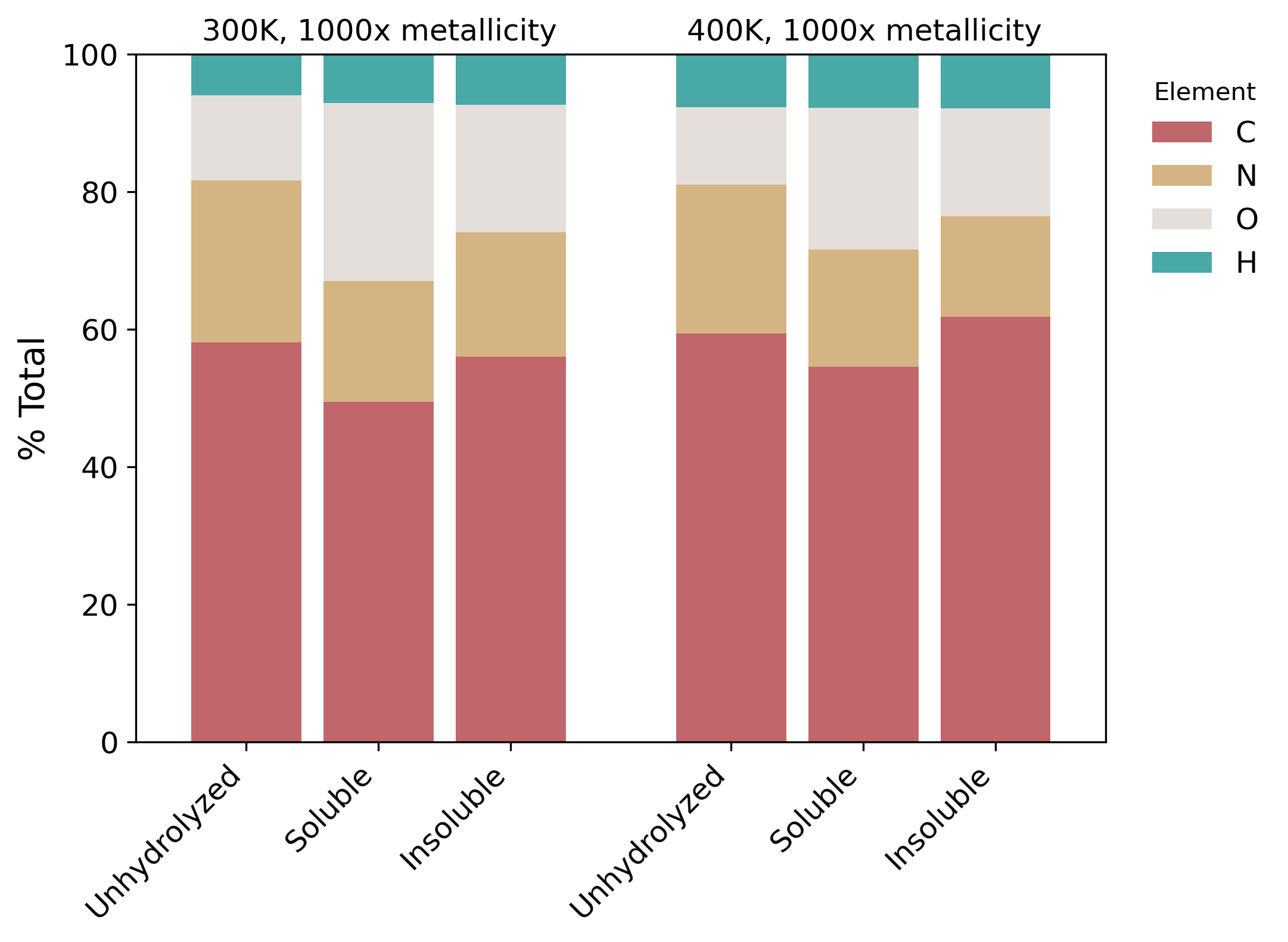}
    \caption{Elemental ratios of unaltered and hydrolyzed haze analogs for two water-rich exoplanets (left: 300 K; right: 400 K). Both sample sets demonstrate increased oxygen after hydrolysis, with a larger increase in oxygen in the soluble fractions. Errors are reported in Table \ref{tab:ea}.}
    \label{fig:ea}
\end{figure}

\begin{table*}[]
\centering
\caption{Elemental ratios and corresponding C/O and C/N ratios for hydrolyzed water-rich exoplanet hazes.}
\begin{tabular}{l l c c c c c c}
    \hline \hline
    Planet & Sample & Carbon wt\% & Hydrogen wt\% & Nitrogen wt\% & Oxygen wt\% & C/O & C/N \\
    \hline
    300 K water-rich & Unhydrolyzed & 58.1 $\pm$ 1.3 & 6.1 $\pm$ 0.4 & 23.5 $\pm$ 2.1 & 12.4 $\pm$ 2.4 & 4.7 $\pm$ 0.9 & 2.5 $\pm$ 0.2 \\
    exoplanet & Soluble & 49.4 $\pm$ 1.1 & 7.2 $\pm$ 0.4 & 17.6 $\pm$ 8.4 & 25.9 $\pm$ 7.2 & 1.9 $\pm$ 0.5 & 2.8 $\pm$ 1.3 \\
     & Insoluble & 56.0 $\pm$ 1.9 & 7.3 $\pm$ 0.4 & 18.1 $\pm$ 6.8 & 18.5 $\pm$ 4.8 & 3.0 $\pm$ 0.8 & 3.1 $\pm$ 1.2 \\
    400 K water-rich & Unhydrolyzed & 59.4 $\pm$ 4.5 & 7.7 $\pm$ 1.1 & 21.6 $\pm$ 2.5 & 11.3 $\pm$ 3.9 & 5.3 $\pm$ 1.9 & 2.7 $\pm$ 0.4 \\
    exoplanet & Soluble & 54.5 $\pm$ 2.2 & 7.8 $\pm$ 0.2 & 17.1 $\pm$ 6.3 & 20.6 $\pm$ 5.0 & 2.6 $\pm$ 0.7 & 3.2 $\pm$ 1.2 \\
     & Insoluble & 61.8 $\pm$ 2.5 & 7.9 $\pm$ 0.5 & 14.6 $\pm$ 4.9 & 15.7 $\pm$ 2.0 & 3.9 $\pm$ 0.5 & 4.2 $\pm$ 1.4 \\
\end{tabular}
\label{tab:ea}
\end{table*}

\begin{figure*}[]
    \centering
    \includegraphics[width=0.95\textwidth]{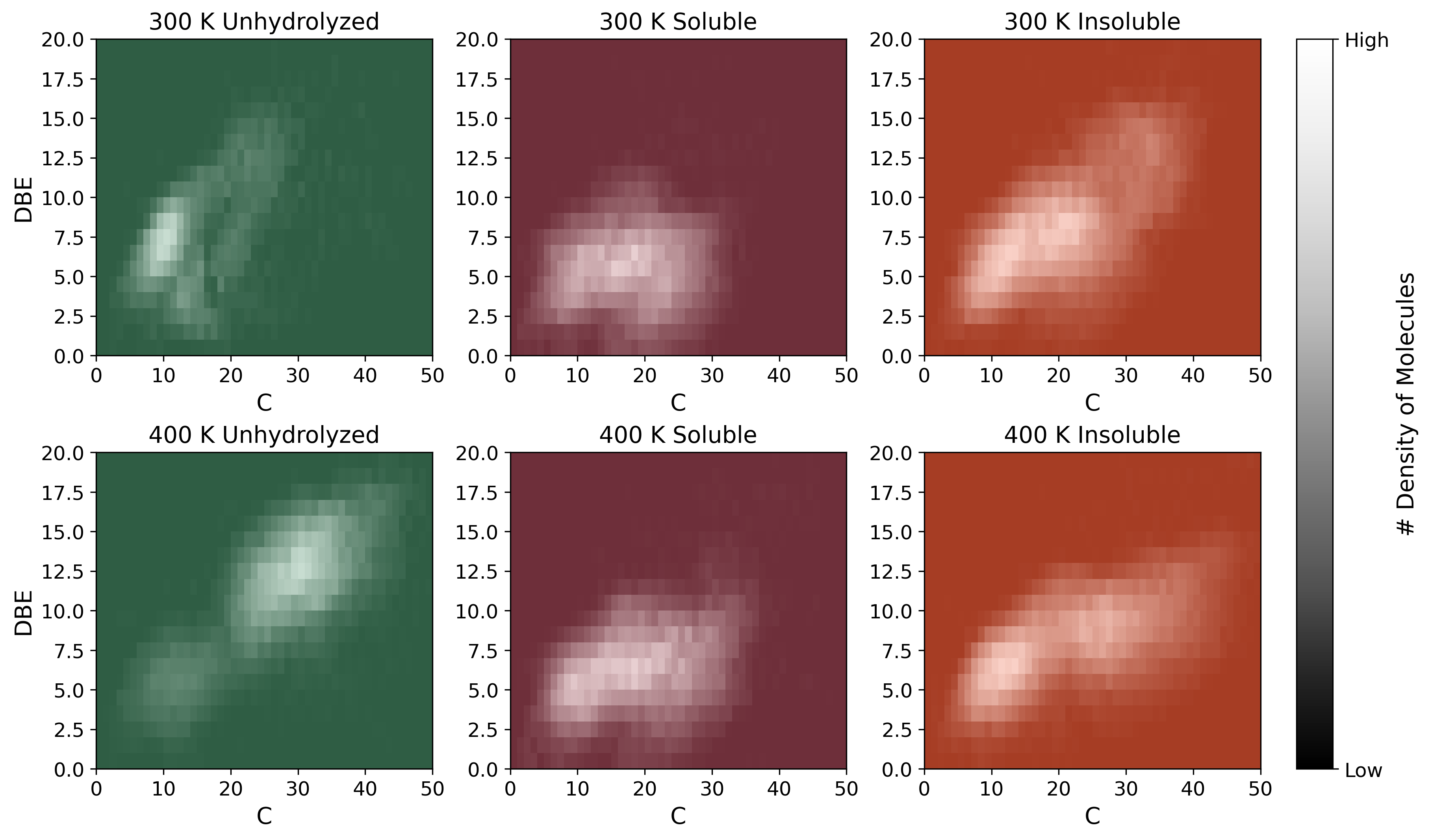}
    \caption{The double bond equivalent (DBE) of each molecule plotted against the number of carbon atoms gives information about the molecular structure. Low and high DBE values for a given carbon number suggest saturated and unsaturated molecules, respectively. The colormap highlights regions with a high number density of molecules, where dark hues are low density and light hues are high density.}
    \label{fig:dbe}
\end{figure*}

We present the elemental ratios of our samples listed in Table \ref{tab:ea} and visualized in Figure \ref{fig:ea}. Notably, the oxygen content in the soluble fraction is higher than that of the unhydrolyzed haze for both planet analogs. The O wt\% for the insoluble fraction lies between the soluble and unhydrolyzed samples, though overlapping errors ($\pm2.0–7.2\%$) warrant caution in the interpretation. This aligns with expectations based on established hydrolysis pathways, where broken-down polymers in the soluble fraction become oxygenated, and highly conjugated molecules in the insoluble fraction do not fragment enough to incorporate as much oxygen. The addition of oxygen comes at the expense of nitrogen and carbon, and distinct chemistries between the planet analogs control the distribution. 

Trends in the elemental analysis are further contextualized using the double bond equivalent (DBE), or degree of unsaturation, plotted in Figure \ref{fig:dbe} and calculated by 
\begin{equation}
    DBE = C+\frac{N}{2}-\frac{H}{2}+1
\end{equation}
The DBE of each molecule can be plotted against the number of carbon atoms to get a sense of how carbon is incorporated into the structure of the molecule \citep{Bae2011,Dong2023}. While the DBE does not uniquely distinguish functional from structural carbon, a lower DBE at a given carbon number reflects a lower degree of unsaturation, whereas a higher DBE reflects a greater degree of unsaturation. Despite stark differences in the original haze analog DBEs, each fraction of the hydrolyzed samples has broadly similar bonding patterns where the soluble fraction occupies a lower DBE space than the insoluble fraction. Combined with observed changes in the elemental analyses, the lower DBE of the soluble fraction is consistent with increased functionalization while the insoluble fraction retains more rings and $\pi$ bonds. We use this information to speculate on the movement of each element throughout hydrolysis.

% 300K: The unhydrolyzed sample contains a large amount of low C, low DBE molecules indicating highly unsaturated, likely aromatic, structures. The tail trailing into the high C, high DBE region continues the trend for larger molecules. The soluble fraction maintains molecule size, if not increasing the amount of C per molecule, but has lower DBE showing more functionalized rather than structural carbon. This means there is more O and/or N filling those structures and that the C wt\% plummets. The insoluble fraction shows the highest C atoms per molecule with high DBE, so oxygen incorporation is more limited and the C wt\% does not decrease as much. There is a clear divide between molecules similar to the unhydrolyzed sample and the insoluble, where the DBE is shifted right and therefore is not quite as saturated as the original. This is supported by the plot vs O where high O low DBE means small fragmented molecules and low O high DBE shows the aromatic backbones remain.

% The 400 K sample begins with higher C and DBE

The 300 K analog is dominated by relatively small molecules with moderate unsaturation, and extensive oxygen incorporation following hydrolysis indicates that the original samples contained abundant heteroatom functional groups susceptible to hydrolysis. Both hydrolyzed fractions become extremely oxygenated, increasing the soluble O wt\% by 13.5\% and even the insoluble by 6.1\%. This is offset primarily by nitrogen for the insoluble fraction, with the N wt\% decreasing by 5.4\%. Hydrolysis efficiently breaks nitrogen-bearing functional groups and releases amines that are either small enough to volatilize or subsequently recombine with oxygen-bearing groups like amides. The C wt\% remains relatively stable due to the carbon frameworks that are difficult to break through hydrolysis. The soluble fraction experiences a similar drop in nitrogen, however there is also significant carbon loss by 8.7\%. Carbon cleaved by hydrolysis in small soluble molecules may be lost as volatile species, or the carbon may join in oxygenated functional groups rather than rigid backbones. The mass of the added oxygen atoms can plummet the C wt\%.

The 400 K analog is primarily made of much larger molecules, evidenced by formulas with large carbon numbers, with a high degree of unsaturation. The larger molecular sizes, higher DBEs, and limited oxygen incorporation together suggest a more condensed molecular framework that is less susceptible to hydrolysis. Oxygenation is present but less intense in the 400 K analog, and the other elemental ratios suggest different bonding tendencies. The insoluble fraction sees a significant N wt\% decrease of 7.0\%, as well as an increase of C wt\% by 2.4\%. Hydrolysis again attacks more nitrogen-bearing functional groups than carbon-bearing functional groups, though this pattern is accentuated by strong structural backbones in the original sample that disfavor fractionation and elevate the C wt\%. The elemental ratios of the soluble fraction behave similarly to the 300 K sample, though the non-recoverable amount of 400 K soluble material obtained in the experiment (see \citealt{Pesciotta2026}) suggests important differences in how the hydrolysis products partition. We speculate that these differences arise from the nature of the original sample. The potentially more condensed molecular framework is less susceptible to extensive hydrolysis, and the heteroatom-containing groups that are cleaved are likely small and sufficiently volatile, resulting in only a small amount of collectible soluble material.

With these interpretations, we also reiterate that the Orbitrap measurements represent the methanol-extractable portion of each sample. Incomplete dissolution in methanol may preferentially exclude larger, less polar, or condensed molecules. The water-insoluble fraction shows the least methanol dissolution, suggesting that the distinction between the soluble and insoluble fractions may be greater than captured here.

Thus, hydrolysis extensively fractionates the smaller, more functionalized 300 K molecular inventory, whereas the larger and more condensed 400 K network is less reactive to hydrolysis and predominantly retains its carbon skeleton. These experiments suggest that hazes produced under different atmospheric conditions respond differently to hydrolysis, leading to differences in oxygen incorporation, molecular reorganization, and partitioning between soluble and insoluble reservoirs.

\subsubsection{Compound Classes}
\begin{figure*}[]
    \centering
    \includegraphics[width=0.9\textwidth]{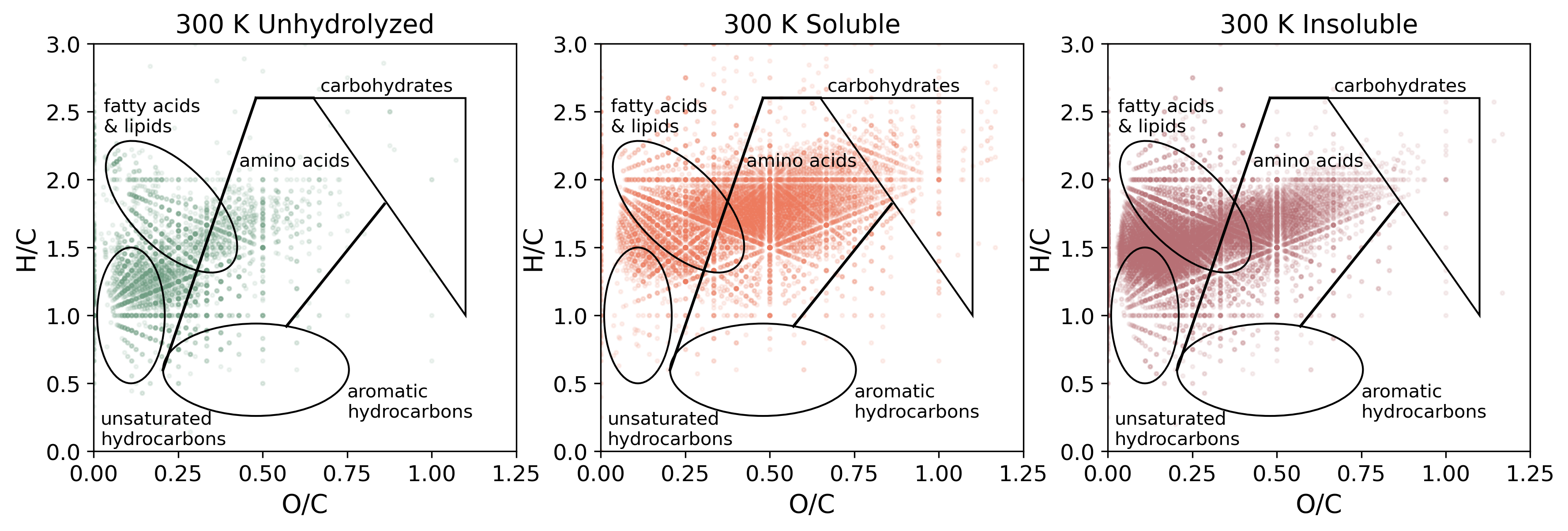}
    \includegraphics[width=0.9\textwidth]{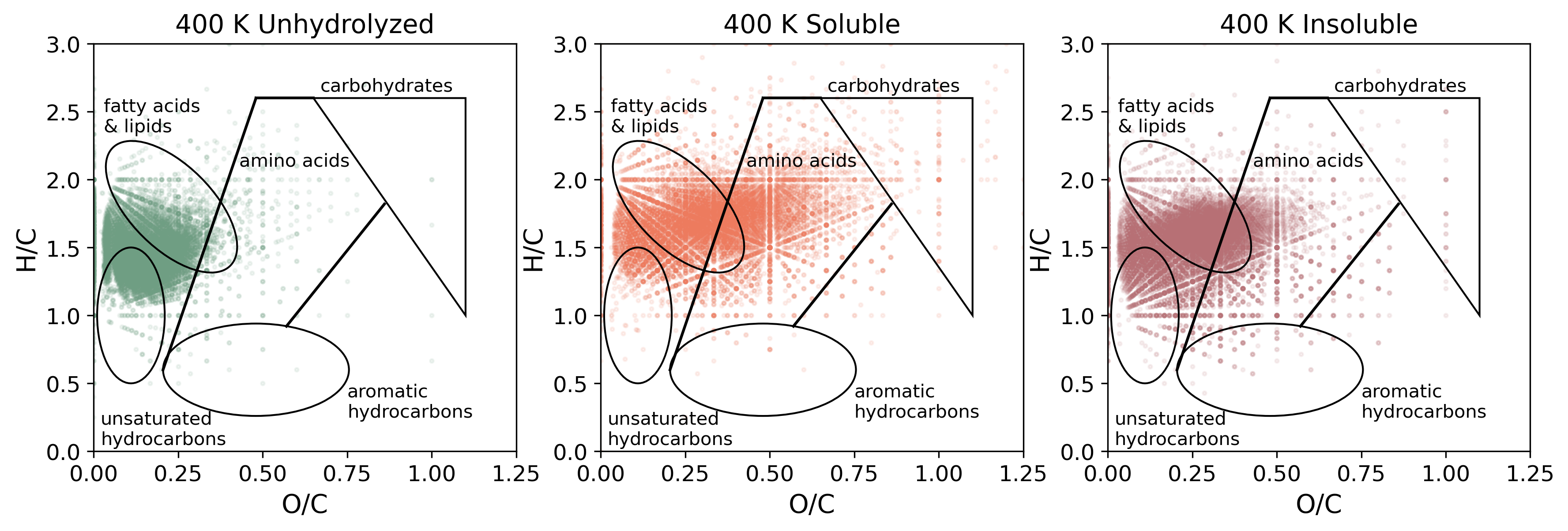}
    \caption{Van Krevelen diagrams of unaltered (left) and hydrolyzed (middle: soluble; right: insoluble) haze analogs for two water-rich exoplanets (top: 300 K; bottom: 400 K). Regions in H/C vs. O/C space are commonly attributed to broad classes of compounds, though these assignments can be less reliable for N-rich samples.}
    \label{fig:vk}
\end{figure*}

Another method to visualize the bulk chemistry in haze analogs is using van Krevelen diagrams (Fig. \ref{fig:vk}). Van Krevelen diagrams plot the molecular H/C vs. O/C ratios to separate complex data into regions of different compound classes including unsaturated and aromatic hydrocarbons, amino acids, fatty acids/lipids, and carbohydrates \citep{Ruf2018}. Distinct diagonal, vertical, and horizontal lines are also visible in this representation, denoting specific chemical pathways associated with hydrogenation, oxidation, and other repeating groups.

Patterns in the van Krevelen diagrams agree with elemental analysis. The average H/C ratio is lower for the original 300 K analog, consistent with the lack of H$_2$ in the original gas mixture and the lower H wt\% in the haze analog. The H/C ratios trend slightly higher for the soluble fraction than the insoluble fraction. The O/C ratio also increases after hydrolysis, confirming oxygenation, and the soluble samples exhibit a higher concentration of formulas at higher O/C ratios than the insoluble samples. The expanded chemical space supports evidence of increased functional group diversity, and clear directional lines are consistent with several reaction pathways that result in repeating chemical families.

Visualized in van Krevelen space, we can place the molecular H/C and O/C ratios in the context of common classes of organic compounds. It is important to note that assignment to a particular region does not necessarily indicate the presence of those compounds, which we would need to confirm through structural analysis, but instead reflects similarities in the elemental composition of C, H, and O. Such compositional similarities may be associated with analogous behavior, including solubility, polarity, and chemical reactivity \citep{Cooper2022}. The unhydrolyzed samples predominantly sit in the unsaturated hydrocarbon and fatty acids and lipids regions. All hydrolyzed samples exhibit large amounts of molecules in the amino acid region and even the carbohydrate region, with the soluble fraction having the most density. The insoluble fractions extend furthest into the aromatic hydrocarbon region.

\subsubsection{Functional Group Constraints}
Thus far, we have only speculated on the functional groups present in these haze analogs based on patterns in the molecular formulas. Our companion paper \citep{Pesciotta2026} presents infrared spectra of the samples and assigns functional groups to observed spectral features, finding changes in OH, NH, CN, CO, and CH-bearing functional groups after hydrolysis (see Figure 2 and Table 4 therein). Together, mass spectrometry and infrared spectroscopy provide stronger constraints on the structure of these molecules, in addition to establishing ground truth for the chemistry in water-altered hazes should they be detected in a planetary atmosphere. We source the functional group spectral modes and associated compounds from \cite{Silverstein2017}.

The morphology of the 2500--3500 cm$^{-1}$ O-H stretch suggests a shift from intra- to intermolecular hydrogen bonding. Hydrolysis splits covalent bonds and inserts oxygen from H$_2$O, creating polar groups that are capable of participating in intermolecular interactions. Because the intermolecular bonds inferred from FTIR do not survive ionization for mass spectrometry, we do not expect to see large carbon numbers in Figure \ref{fig:dbe}. However, the low DBE values across hydrolyzed samples are consistent with the high degree of functionalization created by the OH functional groups responsible for intermolecular bonding. This interpretation is also reflected in the high O wt\% and corresponding decreases in CHN wt\%.

Various nitrogen-bearing spectral features present in the original sample persist in the insoluble fraction, while the soluble fraction loses the broad $\sim$2900 cm$^{-1}$ feature and introduces a distinct free N-H feature at 3413 cm$^{-1}$. This indicates distinct nitrogen speciation between the soluble and insoluble fractions, despite similar N wt\%. One possible interpretation is that water-soluble amine salts dissolve to release free amines. Since this change in bonding environment would not be distinguishable with elemental ratios or DBE, the FTIR features provide the only constraint on this movement of nitrogen.

Nitriles (C$\equiv$N, 2242 cm$^{-1}$) are present in the unhydrolyzed sample and produce amides, carboxylic acids, and amines upon hydrolysis. Indeed, we see strong C=O stretching ($\sim$1630 cm$^{-1}$) characteristic of amides in the soluble fraction as well as several carboxylic acid features. This process is underscored by the high O wt\% increase and subsequent C wt\% decrease, as each amide and carboxylic acid adds one or two oxygen atoms to the molecule. The transformation also decreases the DBE from two in a nitrile to one for either product group, agreeing with the relatively low DBE for the carbon count seen in the soluble fraction.

Interestingly, the soluble fraction exhibits a feature centered around $\sim$2140 cm$^{-1}$ that is consistent with a carbodiimide (N=C=N) or other cumulative double bond stretch. If assigned to carbodiimides, this feature is particularly interesting in a prebiotic context because the compounds can act as coupling reagents that facilitate condensation reactions, including peptide bond formation. Their presence is somewhat unexpected, though, since carbodiimides form through dehydration reactions instead of hydrolysis.

While we have focused primarily on the effects of hydrolysis, the haze analogs likely undergo dehydration reactions facilitated by the drying process required for our analysis methods (see Sections \ref{sec:pbm} and \ref{sec:discuss} for discussions on the implications of dehydration reactions and wet-dry cycling). As such, we can infer that the dehydration reactant species for carbodiimides, dialkylureas, were present in the soluble sample prior to drying. Ureas are plausible products of haze hydrolysis, stemming from the breakdown of nitriles and amides into amines and carbonyls that recombine into ureas in aqueous solutions. Urea has been detected by GC/MS in a Titan haze hydrolysis experiment \citep{Raulin2007} and by GC/MS/MS in a non-hydrolyzed haze analog synthesized from a CH$_4$/CO/N$_2$ gas mixture \citep{Sebree2018}, though it was not detected in the Orbitrap analysis of the oxygen-bearing samples from \cite{Moran2020}. 

The infrared fingerprint region is home to many C-H spectral features that provide clues about the carbon bonding structure from their complex vibrational modes and relation to the rest of the molecule. All samples display modes aligning with aromatic and aliphatic carbon, but importantly, the unhydrolyzed sample exhibits a unique 704 cm$^{-1}$ feature. This low frequency feature corresponds to rocking or out-of-plane bending often seen in minimally substituted rings, where frequency would increase with increased ring substitution. This feature disappears after hydrolysis in favor of a higher frequency 1340 cm$^{-1}$ C-H twisting mode in the soluble fraction, associated with methylene chains and moderated by the surrounding heteroatoms. This agrees with the lower degree of unsaturation and evidence of functionalized carbon in the soluble fraction. 

Lastly, the overall absorption behavior is controlled by bulk chemistry, especially trends in molecular conjugation and complexity. The unhydrolyzed and insoluble samples are moderately absorbing with modest flattening of spectral features in exoplanet atmospheric spectroscopy, while the soluble sample exhibits much larger imaginary refractive indices that have substantial effects on spectral feature amplitude. The broadband absorption seen in the soluble fraction may stem from conjugated systems and heteroatoms, blurring features together and creating continuum absorptions in the spectra.

\subsection{Identification of Prebiotic Molecules}
\label{sec:pbm}

Table \ref{tab:pmsum} summarizes the molecular formulas detected in the samples, as identified by \textit{idmol}, that may correspond to prebiotic molecules. We group the formulas into broad prebiotic compound classes, including derivatives, to illustrate how the chemical diversity changes across samples. Tables \ref{tab:300K_sol}–\ref{tab:400K_insol} in the appendix present the specific molecular formulas and their prebiotic relevance. We only list molecules that were created or destroyed after hydrolysis; for a full list of molecules identified in the original samples, see \cite{Moran2020}. For each formula, the appendix tables report measured mass (m/z), the detection mode the formula is detected in, the formula (C$_w$H$_x$N$_y$O$_z$), potential molecular assignments, and their respective prebiotic relevance. 

Across both formation conditions and hydrolysis states, many new formulas are detected including amino acids, sugars, and their derivatives. There is an exceptional increase in the number of formulas assigned to sugars, compared to originally only detecting the formula for the simplest monosaccharide glyceraldehyde. The list of destroyed molecules, where the corresponding molecular formula is no longer detected, is much shorter, with notable losses including glyceraldehyde and the nucleotide base adenine. We emphasize that high-resolution mass spectrometry cannot determine molecular structure, and that there are many formulas that match such high mass molecules, so these results primarily serve to demonstrate potential abiotic pathways towards prebiotic molecules. 

While hydrolysis is often considered degradative since it cleaves covalent bonds, molecule production seems to dominate over destruction in these samples. This may be due to dehydration-driven condensation reactions that occur during the drying process, though the significant oxygen incorporation demonstrated by elemental analysis (Section \ref{sec:ea}) implies that these processes only contribute minimally to the overall chemical composition. Instead, hydrolysis likely generates reactive intermediates that promote molecular reorganization. Many identified prebiotic formulas are common aqueous transformation products, including oxidation and hydroxylation. 

We detect four unique general formulas for pentose sugars, including ribose and other 5-carbon carbohydrates, across all of the hydrolyzed samples. Pentose sugars are highly unstable, making it difficult to accumulate under aqueous conditions. Sugar alcohols (reduced) and sugar acids (oxidized), however, are more stable than their parent pentose \citep{Zubay2001,VazquezSalazar2026}. The co-detection of pentose sugar formulas with several of their derivatives provides compelling evidence for active carbohydrate chemistry and continuous replenishment that may accumulate ribose for use in nucleoside and, eventually, RNA synthesis \citep{Larralde1995,Powner2009}. We also detect the formula for deoxyribose across all samples, a precursor for the more stable nucleic acid DNA. This chemical inventory demonstrates haze hydrolysis as a productive abiotic formation pathway for the building blocks of life.

While several new sugar formulas are produced, haze hydrolysis may not be conducive to nucleotide base synthesis. No new nucleobases are produced in any of the samples, and adenine is destroyed in the 300 K analog. Adenine is susceptible to hydrolysis, releasing ammonia and converting into hypoxanthine. Guanine, though, survives hydrolysis in both the 300 and 400 K analogs. We also find the molecular formula for dihydroorotic acid, a precursor for pyrimidine nucleobases like uracil, thymine, and cytosine. Continuous wet-dry cycling may be necessary to stimulate nucleobase formation \citep{Damer2020,Pearce2024a}; a mechanism for sufficient nucleobase accumulation to form nucleosides in these environments remains an outstanding issue.

Both the original samples and hydrolyzed samples contain dozens of amino acids, their derivatives, and even dipeptides. Hydrolysis introduces five new potential dipeptides across the samples. The diversity of amino compounds is promising for their catalytic and stabilizing role in prebiotic evolution. Together, the presence of more complex sugars, nucleobase derivatives, and amino compounds demonstrates a greater prebiotic network stemming from the post-processing of photochemical hazes as compared to their initial synthesis.

The original haze analogs provide the first evidence for sugar production without the presence of liquid water \citep{Moran2020}, demonstrating that prebiotic molecules can form directly from atmospheric photochemistry. This challenges the prevailing theory that prebiotic material originates from external delivery by impactors or production in hydrothermal systems. The work performed here further challenges this theory. While we do introduce liquid water, the production of formulas consistent with several monosaccharides and dipeptides in these experiments suggests that material from atmospheres alone can produce increasingly complex prebiotic molecules.

Comparisons with previous haze analog studies illustrate that oxygen availability is a key factor governing molecular complexity. Titan-like hazes, synthesized from N$_2$ and CH$_4$, only form CHN-bearing organics. Subsequent processing is required to introduce oxygenated amino acids and nucleobases into these oxygen-poor haze analogs. Hydrolysis provides one such pathway \citep{Khare1986,McDonald1994,Neish2010,Poch2012}, and other mechanisms, such as ion implantation \citep{Matuszewski2026}, warrant further laboratory investigation. In contrast, haze analogs synthesized from oxygen-bearing gas mixtures are able to produce oxygenated prebiotic molecules, as demonstrated in early-Earth experiments with CO$_2$ \citep{Pearce2024a}; the addition of trace H$_2$S can further expand this chemistry to include sulfur-bearing prebiotic molecules \citep{Reed2025}. Likewise, the more oxygenated sub-Neptune gas mixtures are the first to show molecular formulas consistent with sugars directly through atmospheric photochemistry \citep{Moran2020}. These comparisons place hydrolyzed sub-Neptune haze analogs within a broader progression of increasingly oxygenated prebiotic environments that produce more diverse molecular inventories.

% Diaminobutyric acid and 3-Ureidopropionic acid: found in meteorite samples

% c5h9no3 and c5h11no3 and c7h14n2o3 as formyllysine and dopa: oxidation/hydroxylation

% dihydroxyisovalerate: both meteorite and oxidation

% methylglutamate is an intermediate in methane metabolism

% not just seeing new backbones but secondary functionalization of amino acids

% so the bile acids... likely not actually bile acid, duh. they have 4 fused rings which is kinda crazy, though our ftir show a lot more aromatic rings. more likely it's a long chain polyfunctional organic that happens to have the same mass

% formylkynurenine is breakdown product of tryptophan?

\begin{table*}[]
    \caption{Summary of the molecular formulas associated with each compound class that are created (+) or destroyed (–) after hydrolysis. For formulas that can be associated with multiple compound classes, the formula is counted once for each applicable class. These counts summarize potential chemical diversity rather than the number of uniquely identified compounds. Full tables with a list of the molecules that were created and destroyed are included in the appendix.}
    \centering
    \begin{tabular}{l c c c c}
        \hline\hline
        \textbf{Compound class} & \textbf{300 K Sol.} & \textbf{300 K Insol.} & \textbf{400 K Sol.} & \textbf{400 K Insol.}  \\
        \hline
        Sugars \& derivatives & +8/–1 & +10/–1 & +11/–0 & +11/–0 \\
        Nucleobases \& derivatives & +1/–1 & +1/–1 & +1/–0 & +1/–0 \\
        Amino compounds & +28/–6 & +36/–1 & +34/–1 & +37/–0 \\
        Total & +38/–8 & +47/–3 & +45/–1 & +49/–0 \\
    \end{tabular}
    
    \label{tab:pmsum}
\end{table*}

\section{Discussion}
\label{sec:discuss}

The habitability of sub-Neptunes is poorly constrained due to the sweeping degeneracies that exist for their interiors and atmospheres. The proposed water environments vary widely depending on temperature, atmospheric structure, and metallicity. Warm sub-Neptunes can sustain large abundances of atmospheric water vapor at high altitudes where it actively participates in photochemistry \citep{Beatty2024}. At sufficient pressures and temperatures deep in the atmosphere, water may exist in high-pressure states or as a miscible gas with the dominant H$_2$ envelope \citep{Piaulet2024,Benneke2024,Luu2024}. In cooler atmospheres, water can condense out to form clouds. On the coldest sub-Neptunes, these clouds may create a cold trap that prevents water vapor from mixing upward into the upper atmosphere, sequestering water at depths where the pressures and temperatures can support liquid water \citep{Madhusudhan2023,Huang2024,Hu2025}. These regimes delineate where hazes must subsist to react with liquid water, and each regime has its own set of complications for life formation and sustainability.

Photochemical hazes are most likely to interact with liquid water in condensate clouds or oceans. While hazes primarily form in upper atmospheres, the particles can traverse to cloud or surface altitudes through wet or dry deposition \citep{Yu2021,Lavvas2026}. At altitudes where water condenses, hazes can serve as CCN \citep{Mang2022,Lavvas2026}, with water-rich haze analogs showing particularly favorable properties for CCN \citep{Yu2021,Pesciotta2026}, and chemical reactions can occur at this cloud-particle interface \citep{Maillard2023}. Quantifying the amount of hydrolyzed haze in the atmosphere would therefore require coupling microphysical models and aerosol residence times to the soluble and insoluble partitioning measured here. Clouds can be harsh environments for microorganisms due to extreme conditions and nutrient limitations, though they may be safe havens on planets like Venus where high temperatures, pressures, and acidity pose astrobiological challenges \citep{Kotsyurbenko2021}. Similar considerations should be explored for cloudy conditions on sub-Neptunes.

Deeper down, hazes may deposit onto ocean surfaces where they can remain at the interface, become suspended in the ocean, or sink depending on properties like relative density, particle size, and ocean dynamics. While \cite{Yu2021} provide important laboratory constraints on the properties that control settling, quantifying the amount of haze ultimately reaching an ocean surface will require coupled aerosol transport and microphysical modeling. Importantly, the pressures at these ocean surfaces ($>$100 bar; \citealt{Oger2010,Rigby2024}) do not preclude biologically-relevant chemistry. They may be comparable to the pressures in hydrothermal vent systems on Earth, where life persists down to our ocean's deepest point at 1000 bar \citep{Rothschild2001,Merino2019}. Should haze material make it to the seafloor (up to $10^5$ bar; \citealt{Rigby2024}), hazes may contribute an additional source of material for prebiotic chemistry at fluid-rock interfaces. Future work could explore the catalytic properties of fluid-rock-haze systems.

Without the presence of a solid surface, any wet-dry cycling would have to occur as condensation and evaporation of cloud particles. As discussed in Section \ref{sec:pbm}, this cycling may be important for expanding and stabilizing the chemical network. The re-releasing of CCN into the atmosphere is a common process on Earth known as regeneration, often leaving behind larger particles that are even better CCN candidates \citep{Arieli2025}. Though most wet-dry cycling studies focus on surface features like hot springs, ponds, and shorelines, this example from Earth suggests that cycling could occur solely in atmospheres. 

As hydrolysis partitions haze material into chemically distinct soluble and insoluble reservoirs, these fractions likely experience different transport processes based on their diverging physical and chemical properties. In atmospheres, insoluble particles may remain suspended or sediment into deeper atmospheric layers. Upon reaching surface waters, they may remain suspended or settle toward the seafloor, depending on their physical properties and local dynamics. Analogous partitioning occurs in Earth's oceans, where algae and organic material accumulates in surface films \citep{Garrett1967} while denser particulates like marine snow and sediments settle \citep{Buesseler2009}. Dissolved products may remain in solution, undergo aqueous cycling, or evaporate back into the gas phase where they can again participate in cloud formation and atmospheric chemistry. Future work could explore how the physical and chemical properties of aqueously-altered hazes influence their long-term evolution in planetary environments.

There are many challenges in directly detecting hydrolyzed hazes on exoplanets with current observational capabilities. Upcoming telescopes like Roman and HWO are aiming to see down to exoplanet surfaces, including oceans \citep{Hu2025}, though sparse or sinking particles would not be detectable. A thick, uniform floating layer of haze on the ocean surface would likely be the only way to detect an anomalous albedo. Detecting hydrolyzed hazes in atmospheres is more feasible but still challenging. Model reflectance and transmission spectra of high-altitude hydrolyzed hazes show distinct spectral behaviors that may be observable under opportune conditions \citep{Pesciotta2026}. This requires a sufficient haze number density, a haze layer at observable altitudes, and excellent signal-to-noise. Even with the challenges in detecting distinct haze chemistries, analog haze materials can tell us what planetary conditions support haze formation and allow us to surmise the prebiotic chemistry occurring in these environments without in situ measurements. Should spectral evidence for aqueous processing of photochemical hazes be observed on a water-rich exoplanet, we could infer a reservoir of complex organic material on the planet.

Although this discussion focuses on sub-Neptunes due to their observational advantages over super-Earths and terrestrial planets, many of the same haze-water interactions may also occur on rocky planets with photochemically-active atmospheres. Organic hazes may similarly encounter clouds, surface waters, or hydrothermal systems, but rocky planets may provide more favorable opportunities for prebiotic chemistry through lower-pressure aqueous environments and surface wet-dry cycling. Observations of these worlds may resemble Titan-like exoplanets, where reflected light could distinguish among surface albedo features such as lakes and seas, ice, and deposited haze material.

\section{Conclusions}
This study analyzes the chemical evolution of two sub-Neptune haze analogs after exposure to water to explore potential prebiotic processes at the interfaces of organic material and water reservoirs. High-resolution mass spectrometry measurements of water-altered photochemical hazes detect thousands of individual CHNO molecules. The two studied planet analogs are chemically distinct, resulting from the different initial starting conditions, though hydrolysis drives their properties to become more similar. Elemental and structural analysis demonstrate that oxygen is efficiently incorporated into the samples, significantly in both soluble fractions and moderately in the insoluble fractions. Accordingly, the soluble fraction contains more functionalized carbon compared to structural carbon in the insoluble fraction. Van Krevelen diagrams also show the molecules extending far into O/C vs. H/C space, indicating increased chemical complexity and a diversity of compounds. 

When we compare the identified molecular formulas to those of important prebiotic molecules, we find evidence for many new molecules including monosaccharides and their derivatives, and few molecules are destroyed. Future work should seek to confirm the presence of amino acids, nucleobases, and sugars with additional mass spectrometry techniques. These experiments continue to suggest that a rich prebiotic network can form in atmospheres and surfaces alone, and laboratory work will be imperative for linking observations to prebiotic chemistry in hazy exoplanet atmospheres.

\section{Acknowledgments}
C.P. is supported by the NASA Future Investigators in NASA Earth and Space Science and Technology grant 80NSSC25K0305. Portions of this work was supported by the NASA Exoplanets Research Program grant NNX16AB45G and the NSF Astronomy and Astrophysics Grant Program grant 2206245.

\bibliography{mybib}

\clearpage
\appendix
\section{Mass Defect Analysis}
\label{sec:md}
\begin{figure}[h]
    \centering
    \includegraphics[width=0.9\textwidth]{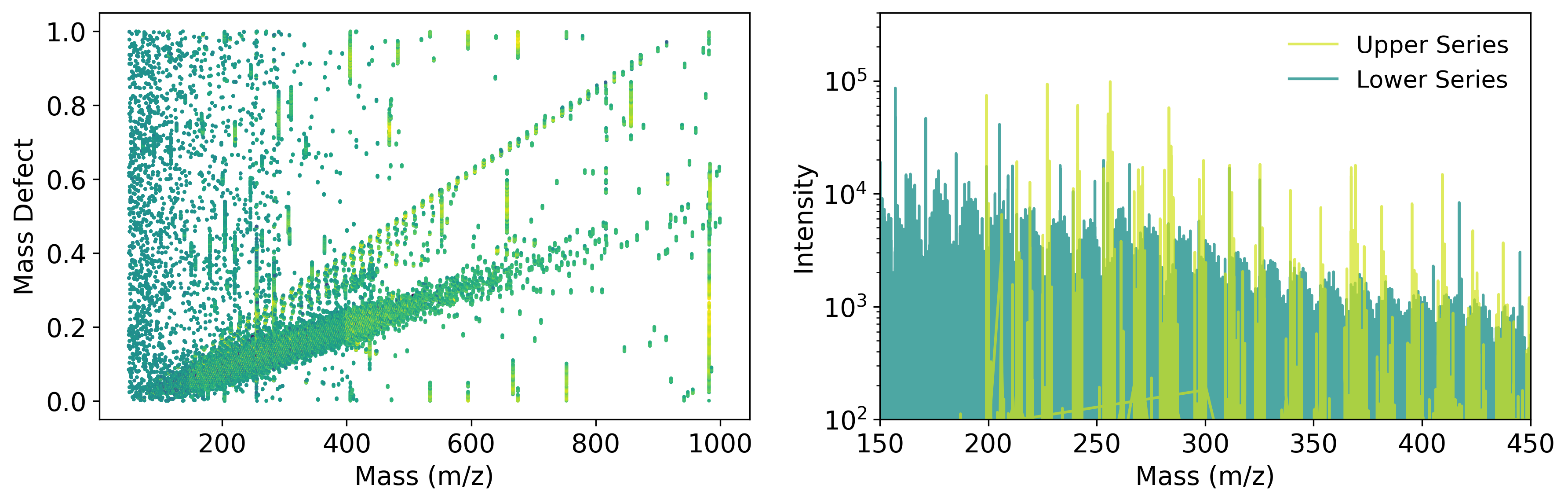}
    \caption{Left panel: Mass defect of the 300 K unhydrolyzed haze analog. Two clouds of repeating chemical families emerge. Right panel: Mass spectrum separating contributions from the two primary mass defect clouds.}
    \label{fig:md}
\end{figure}

High-resolution mass spectrometry allows us to plot the mass defect of the peaks, calculated by subtracting each peak's nominal mass from the exact mass. This separates distinct chemical families into diagonal lines. The slope of these lines can reveal what homologous group is repeating. In the left panel of Figure \ref{fig:md}, we present the mass defect of the 300 K unhydrolyzed positive mode spectra to demonstrate the cleaning procedure and expand on the unique spectral shape of this sample.

Clear structures emerge from the data. The widespread diffuse cloud spanning mass defects from 0 to 1 and $<$400 m/z are spectral noise, and many of these data points disappear with background subtraction. The nearly vertical stripes, notably disconnected from the other structures, are contaminants or data processing artifacts. These are removed to maintain compatibility with \textit{idmol}, which uses chain growth mechanisms to make molecular assignments.

The dense cloud spanning mass defects from 0 to 0.5 is the main organic inventory. While a detailed analysis of the slope is beyond the scope of this work, the trend generally matches a repeated C$_x$H$_y$N$_z$ group. There is a separate, distinct pattern with mass defects from 0.2 to 1. The data start at low masses and branch off of the main cloud, suggesting that this is a real component of the sample that begins a new homologous series. This slope more closely resembles a CH$_2$ repeating family. We can isolate the peaks responsible for these two clouds, called `Upper Series' and `Lower Series', and examine their contributions to the mass spectrum. The right panel in Figure \ref{fig:md} shows that the expected bell-like structure evidently separates from the sharp repeating peaks. Future work may aim to identify the homologous series and compare them to the other haze analogs to decipher why the sharp peaks are unique to this sample.

\clearpage
\section{Additional Mass Spectra}
\label{sec:addms}
\begin{figure}[h]
    \centering
    \includegraphics[width=0.9\textwidth]{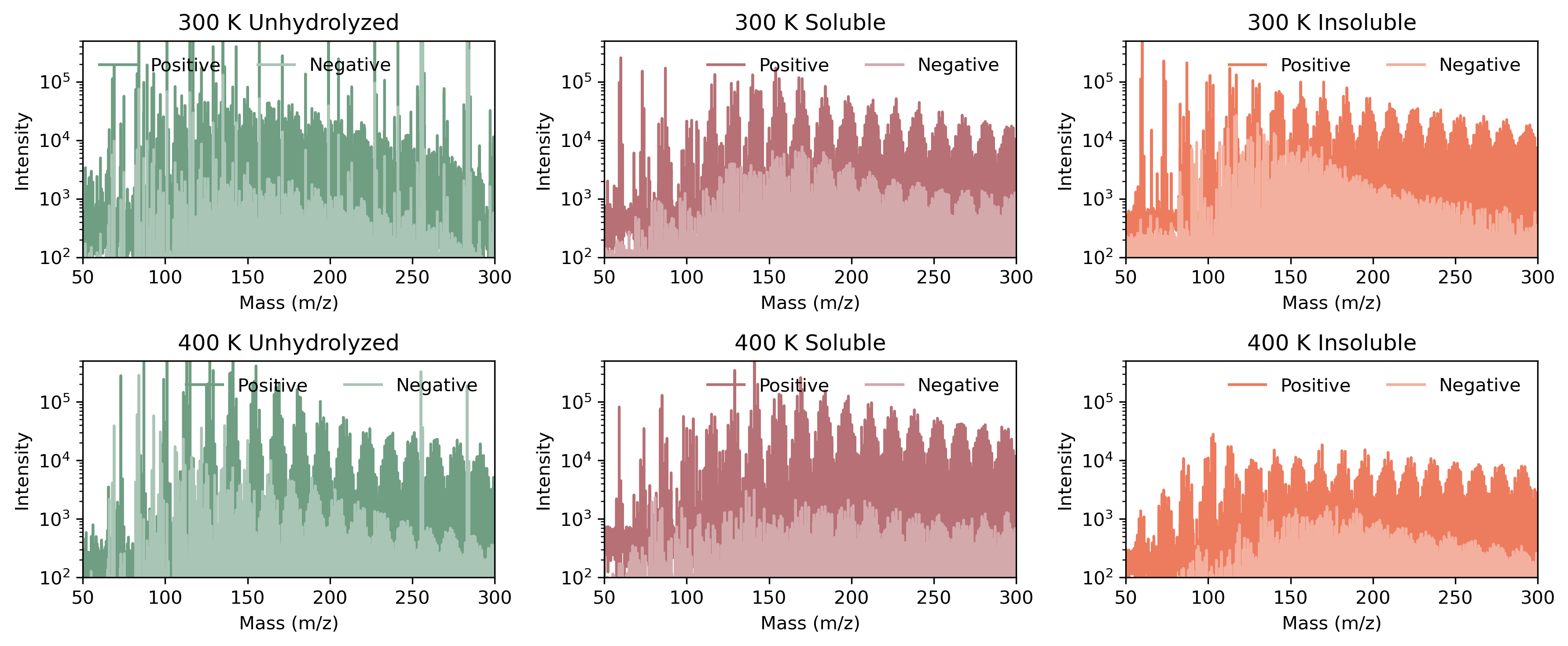}
    \includegraphics[width=0.9\textwidth]{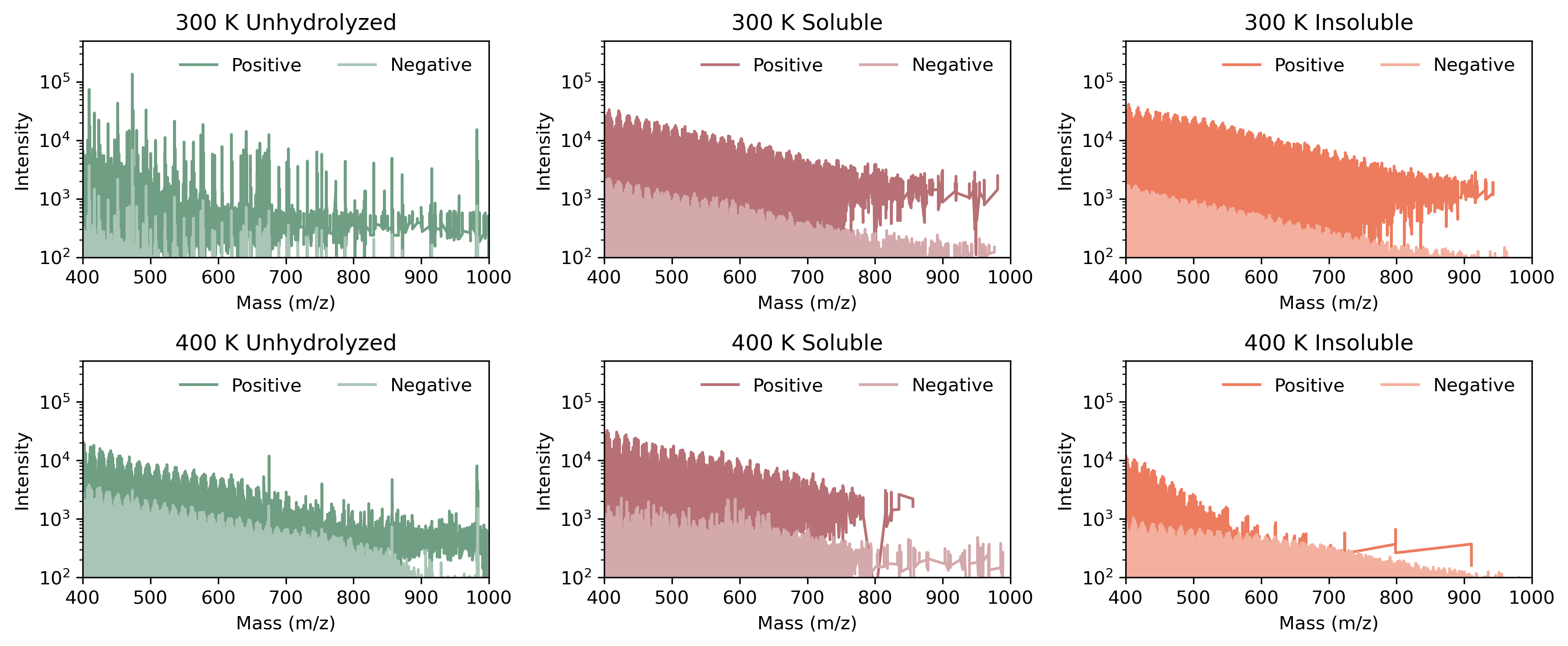}
    \caption{Mass spectra for the additional mass bins. The top two rows present the low mass range (50–300 m/z) for the 300 (top) and 400 K (bottom) planet analogs, and the bottom two rows present the high mass range (400–1000 m/z). For each mass bin, we plot the positive and negative mode measurements of the unhydrolyzed, soluble, and insoluble samples.}
    \label{fig:ms_appendix}
\end{figure}

Mass spectra for the low (50–300 m/z) and high mass (400–1000 m/z) bins are shown here for completeness. These spectra were processed using the same cleaning and \textit{idmol} analysis procedures described in Section \ref{sec:methods}. The spectral shapes are consistent with those described for the 150–300 m/z bins in Section \ref{sec:ms}. Elemental analysis, double bond equivalent, compound class, and prebiotic molecule analyses in the main text all include peaks in these mass bins.

\clearpage
\section{Identification of Prebiotic Molecules}

\begin{longtable*}{c c c l l}
\caption{Molecular formulas created and destroyed after hydrolysis in the soluble fraction of the 300 K water-rich haze.} \\
\hline \hline
Mass (m/z) $\pm$ $\Delta$ppm & Mode & Formula & Potential Molecule & Relevance \\
\hline
\endfirsthead

\multicolumn{5}{c}%
{{\bfseries \tablename\ \thetable{} -- continued from previous page}} \\
\hline
Mass (m/z) $\pm$ $\Delta$ppm & Mode & Formula & Potential Molecule & Relevance \\
\hline
\endhead

\hline \multicolumn{5}{r}{{Continued on next page}} \\
\endfoot

\hline
\endlastfoot
    
\multicolumn{1}{l}{\textit{\textbf{Created}}} \\
\hline
103.0633 $\pm$ 8.7 & - & C$_{4}$H$_{9}$NO$_{2}$ & aminobutyric acid & non-proteinogenic amino acid \\
 & & & dimethylglycine & amino acid derivative \\
117.079 $\pm$ 3.4 & - & C$_{5}$H$_{11}$NO$_{2}$ & valine & biological amino acid \\
 & & & trimethylglycine & amino acid derivative \\
118.0266 $\pm$ 9.5 & - & C$_{4}$H$_{6}$O$_{4}$ & succinic acid & dicarboxylic acid \\
118.0742 $\pm$ 2.7 & + & C$_{4}$H$_{10}$N$_{2}$O$_{2}$ & diaminobutyric acid & non-proteinogenic amino acid \\
 & & & $\beta$-methylamino-L-alanine & non-proteinogenic amino acid \\
129.079 $\pm$ 4.7 & - & C$_{6}$H$_{11}$NO$_{2}$ & cycloleucine & non-proteinogenic amino acid \\
131.0582 $\pm$ 1.2 & - & C$_{5}$H$_{9}$NO$_{3}$ & aminolevulinic acid & non-proteinogenic amino acid \\
 & & & hydroxyproline & amino acid derivative \\
132.0535 $\pm$ 1.3 & - & C$_{4}$H$_{8}$N$_{2}$O$_{3}$ & 3-ureidopropionic acid & amino acid derivative \\
 & & & glycylglycine & dipeptide \\
 & & & asparagine & biological amino acid \\
133.0375 $\pm$ 6.6 & - & C$_{4}$H$_{7}$NO$_{4}$ & iminodiacetic acid & amino acid derivative \\
 & & & aspartic acid & biological amino acid \\
133.0739 $\pm$ 2.4 & - & C$_{5}$H$_{11}$NO$_{3}$ & hydroxyaminovaleric acid & non-proteinogenic amino acid \\
134.0579 $\pm$ 4.4 & - & C$_{5}$H$_{10}$O$_{4}$ & dihydroxyisovalerate & amino acid metabolite \\
 & & & deoxyribose & monosaccharide \\
136.0372 $\pm$ 2.7 & - & C$_{4}$H$_{8}$O$_{5}$ & threonic acid & sugar acid \\
 & & & erythronic acid & sugar acid \\
146.0691 $\pm$ 1.5 & + & C$_{5}$H$_{10}$N$_{2}$O$_{3}$ & $\beta$-ureidoisobutyric acid & amino acid metabolite \\
 & & & glutamine & biological amino acid \\
147.0532 $\pm$ 3.8 & - & C$_{5}$H$_{9}$NO$_{4}$ & o-cetylserine & non-proteinogenic amino acid \\
 & & & glutamic acid & biological amino acid \\
150.0528 $\pm$ 3.0 & - & C$_{5}$H$_{10}$O$_{5}$ & ribose & monosaccharide \\
 & & & pentose sugars & monosaccharides \\
152.0685 $\pm$ 7.3 & + & C$_{5}$H$_{12}$O$_{5}$ & ribitol & sugar alcohol \\
 & & & pentitols & sugar alcohols \\
157.0375 $\pm$ 6.8 & - & C$_{6}$H$_{7}$NO$_{4}$ & 2-aminomuconic acid & non-proteinogenic amino acid \\
158.0328 $\pm$ 1.5 & - & C$_{5}$H$_{6}$N$_{2}$O$_{4}$ & dihydroorotic acid & nucleobase precursor \\
159.0895 $\pm$ 0.4 & - & C$_{7}$H$_{13}$NO$_{3}$ & betonicine & amino acid metabolite \\
 & & & acetylvaline & amino acid metabolite \\
161.0688 $\pm$ 2.6 & - & C$_{6}$H$_{11}$NO$_{4}$ & $\alpha$-aminoadipic acid & non-proteinogenic amino acid \\
 & & & methylglutamate & amino acid metabolite \\
172.096 $\pm$ 0.1 & + & C$_{6}$H$_{12}$N$_{4}$O$_{2}$ & enduracididine & non-proteinogenic amino acid \\
173.0437 $\pm$ 0.6 & - & C$_{5}$H$_{7}$N$_{3}$O$_{4}$ & azaserine & non-proteinogenic amino acid \\
174.1004 $\pm$ 0.1 & + & C$_{7}$H$_{14}$N$_{2}$O$_{3}$ & theanine & non-proteinogenic amino acid \\
 & & & formyllysine & non-proteinogenic amino acid \\
174.1117 $\pm$ 0.5 & + & C$_{6}$H$_{14}$N$_{4}$O$_{2}$ & arginine & biological amino acid \\
175.0845 $\pm$ 1.8 & - & C$_{7}$H$_{13}$NO$_{4}$ & spermidic acid & non-proteinogenic amino acid \\
 & & & valienamine & amino sugar \\
175.0957 $\pm$ 1.0 & - & C$_{6}$H$_{13}$N$_{3}$O$_{3}$ & citrulline & non-proteinogenic amino acid \\
176.0909 $\pm$ 8.5 & + & C$_{5}$H$_{12}$N$_{4}$O$_{3}$ & canavanine & non-proteinogenic amino acid \\
180.0634 $\pm$ 9.2 & - & C$_{6}$H$_{12}$O$_{6}$ & glucose & monosaccharide \\
 & & & hexoses & monosaccharides \\
182.079 $\pm$ 4.7 & + & C$_{6}$H$_{14}$O$_{6}$ & sorbitol & sugar alcohol \\
 & & & hexitols & sugar alcohols \\
188.1161 $\pm$ 0.5 & + & C$_{8}$H$_{16}$N$_{2}$O$_{3}$ & acetyl lysine & amino acid derivative \\
 & & & glycyl leucine & dipeptide \\
 & & & alanyl valine & dipeptide \\
188.1273 $\pm$ 0.9 & + & C$_{7}$H$_{16}$N$_{4}$O$_{2}$ & homoarginine & non-proteinogenic amino acid \\
 & & & methylarginine & amino acid derivative \\
195.0532 $\pm$ 9.2 & - & C$_{9}$H$_{9}$NO$_{4}$ & dopaquinone & amino acid derivative \\
 & & & salicyluric acid & amino acid metabolite \\
196.0583 $\pm$ 4.0 & + & C$_{6}$H$_{12}$O$_{7}$ & gluconic acid & sugar acid \\
 & & & hexonic acids & sugar acids \\
197.0688 $\pm$ 1.7 & - & C$_{9}$H$_{11}$NO$_{4}$ & dihydroxyphenylalanine & amino acid derivative \\
236.0797 $\pm$ 2.0 & +/- & C$_{11}$H$_{12}$N$_{2}$O$_{4}$ & formylkynurenine & amino acid derivative \\
246.1328 $\pm$ 0.1 & - & C$_{9}$H$_{18}$N$_{4}$O$_{4}$ & octopine & amino acid metabolite \\
270.0964 $\pm$ 0.1 & - & C$_{10}$H$_{14}$N$_{4}$O$_{5}$ & aspartyl histidine & dipeptide \\
449.3141 $\pm$ 6.8 & + & C$_{26}$H$_{43}$NO$_{5}$ & glycodeoxycholic acid & bile acid \\
465.309 $\pm$ 6.5 & + & C$_{26}$H$_{43}$NO$_{6}$ & glycocholic acid & bile acid \\
\hline
\multicolumn{1}{l}{\textit{\textbf{Destroyed}}} \\
\hline
%76.0161 $\pm$ 8.4 & - & C$_{2}$H$_{4}$O$_{3}$ & glycolic acid & sugar derivative \\
90.0317 $\pm$ 4.5 & - & C$_{3}$H$_{6}$O$_{3}$ & glyceraldehyde & monosaccharide \\
 & & & lactic acid & sugar derivative \\
%113.0477 $\pm$ 8.1 & - & C$_{5}$H$_{7}$NO$_{2}$ & propargylglycine & non-proteinogenic amino acid \\
%129.0426 $\pm$ 6.3 & - & C$_{5}$H$_{7}$NO$_{3}$ & pyroglutamic acid & non-proteinogenic amino acid \\
% & & & $\beta$-ethynylserine & non-proteinogenic amino acid \\
135.0545 $\pm$ 5.8 & - & C$_{5}$H$_{5}$N$_{5}$ & adenine & nucleotide base \\
137.0477 $\pm$ 5.9 & - & C$_{7}$H$_{7}$NO$_{2}$ & homarine & non-proteinogenic amino acid \\
141.0426 $\pm$ 4.9 & - & C$_{6}$H$_{7}$NO$_{3}$ & 2-aminomuconic semialdehyde & amino acid metabolite \\
153.0426 $\pm$ 2.6 & - & C$_{7}$H$_{7}$NO$_{3}$ & aminosalicylic acids & amino acid derivative \\
 & & & hydroxyanthranilic acid & amino acid metabolite \\
179.0946 $\pm$ 4.5 & - & C$_{10}$H$_{13}$NO$_{2}$ & homophenylalanine & non-proteinogenic amino acid \\
195.0895 $\pm$ 3.4 & - & C$_{10}$H$_{13}$NO$_{3}$ & methyl tyrosine & non-proteinogenic amino acid \\
 & & & homotyrosine & non-proteinogenic amino acid \\
204.0899 $\pm$ 3.0 & - & C$_{11}$H$_{12}$N$_{2}$O$_{2}$ & tryptophan & biological amino acid \\
\label{tab:300K_sol}
\end{longtable*}

\begin{longtable*}{c c c l l}
\caption{Molecular formulas created and destroyed after hydrolysis in the insoluble fraction of the 300 K water-rich haze.} \\
\hline \hline
Mass (m/z) $\pm$ $\Delta$ppm & Mode & Formula & Potential Molecule & Relevance \\
\hline
\endfirsthead

\multicolumn{5}{c}%
{{\bfseries \tablename\ \thetable{} -- continued from previous page}} \\
\hline
Mass (m/z) $\pm$ $\Delta$ppm & Mode & Formula & Potential Molecule & Relevance \\
\hline
\endhead

\hline \multicolumn{5}{r}{{Continued on next page}} \\
\endfoot

\hline
\endlastfoot

\multicolumn{1}{l}{\textit{\textbf{Created}}} \\
\hline
101.0477 $\pm$ 2.8 & - & C$_{4}$H$_{7}$NO$_{2}$ & azetidine-2-carboxylic acid & non-proteinogenic amino acid \\
 & & & vinyl glycine & non-proteinogenic amino acid \\
103.0633 $\pm$ 3.6 & - & C$_{4}$H$_{9}$NO$_{2}$ & aminobutyric acid & non-proteinogenic amino acid \\
 & & & dimethylglycine & amino acid derivative \\
115.0633 $\pm$ 1.6 & - & C$_{5}$H$_{9}$NO$_{2}$ & proline & biological amino acid \\
 & & & allylglycine & amino acid derivative \\
117.079 $\pm$ 2.1 & - & C$_{5}$H$_{11}$NO$_{2}$ & valine & biological amino acid \\
 & & & trimethylglycine & amino acid derivative \\
118.0742 $\pm$ 2.7 & + & C$_{4}$H$_{10}$N$_{2}$O$_{2}$ & diaminobutyric acid & non-proteinogenic amino acid \\
 & & & $\beta$-methylamino-L-alanine & non-proteinogenic amino acid \\
122.0579 $\pm$ 7.8 & - & C$_{4}$H$_{10}$O$_{4}$ & erythritol & sugar alcohol \\
 & & & threitol & sugar alcohol \\
129.079 $\pm$ 1.1 & - & C$_{6}$H$_{11}$NO$_{2}$ & cycloleucine & non-proteinogenic amino acid \\
131.0582 $\pm$ 0.7 & - & C$_{5}$H$_{9}$NO$_{3}$ & aminolevulinic acid & non-proteinogenic amino acid \\
 & & & hydroxyproline & amino acid derivative \\
132.0535 $\pm$ 0.5 & - & C$_{4}$H$_{8}$N$_{2}$O$_{3}$ & 3-ureidopropionic acid & amino acid derivative \\
 & & & glycylglycine & dipeptide \\
 & & & asparagine & biological amino acid \\
132.0899 $\pm$ 2.7 & + & C$_{5}$H$_{12}$N$_{2}$O$_{2}$ & ornithine & non-proteinogenic amino acid \\
133.0375 $\pm$ 0.7 & - & C$_{4}$H$_{7}$NO$_{4}$ & iminodiacetic acid & amino acid derivative \\
 & & & aspartic acid & biological amino acid \\
134.0579 $\pm$ 0.8 & - & C$_{5}$H$_{10}$O$_{4}$ & dihydroxyisovalerate & amino acid metabolite \\
 & & & deoxyribose & monosaccharide \\
143.0946 $\pm$ 1.6 & + & C$_{7}$H$_{13}$NO$_{2}$ & stachydrine & amino acid derivative \\
146.0691 $\pm$ 1.0 & - & C$_{5}$H$_{10}$N$_{2}$O$_{3}$ & $\beta$-ureidoisobutyric acid & amino acid metabolite \\
 & & & glutamine & biological amino acid \\
147.0532 $\pm$ 0.5 & - & C$_{5}$H$_{9}$NO$_{4}$ & o-cetylserine & non-proteinogenic amino acid \\
 & & & glutamic acid & biological amino acid \\
150.0528 $\pm$ 9.5 & - & C$_{5}$H$_{10}$O$_{5}$ & ribose & monosaccharide \\
 & & & pentose sugars & monosaccharides \\
152.0685 $\pm$ 7.3 & + & C$_{5}$H$_{12}$O$_{5}$ & ribitol & sugar alcohol \\
 & & & pentitols & sugar alcohols \\
157.0375 $\pm$ 0.5 & - & C$_{6}$H$_{7}$NO$_{4}$ & 2-aminomuconic acid & non-proteinogenic amino acid \\
157.1103 $\pm$ 0.9 & + & C$_{8}$H$_{15}$NO$_{2}$ & cyclohexylglycine & non-proteinogenic amino acid \\
158.0328 $\pm$ 0.1 & - & C$_{5}$H$_{6}$N$_{2}$O$_{4}$ & dihydroorotic acid & nucleobase precursor \\
159.0895 $\pm$ 0.7 & - & C$_{7}$H$_{13}$NO$_{3}$ & betonicine & amino acid metabolite \\
 & & & acetylvaline & amino acid metabolite \\
161.0688 $\pm$ 0.5 & - & C$_{6}$H$_{11}$NO$_{4}$ & $\alpha$-aminoadipic acid & non-proteinogenic amino acid \\
 & & & methylglutamate & amino acid metabolite \\
164.0685 $\pm$ 8.2 & - & C$_{6}$H$_{12}$O$_{5}$ & fucose & sugar derivative \\
 & & & deoxy hexoses & sugar derivatives \\
166.0477 $\pm$ 9.4 & - & C$_{5}$H$_{10}$O$_{6}$ & ribonic acid & sugar acid \\
 & & & pentose acids & sugar acids \\
172.096 $\pm$ 1.2 & + & C$_{6}$H$_{12}$N$_{4}$O$_{2}$ & enduracididine & non-proteinogenic amino acid \\
173.0437 $\pm$ 0.2 & - & C$_{5}$H$_{7}$N$_{3}$O$_{4}$ & azaserine & non-proteinogenic amino acid \\
174.1004 $\pm$ 0.1 & +/- & C$_{7}$H$_{14}$N$_{2}$O$_{3}$ & theanine & non-proteinogenic amino acid \\
 & & & formyllysine & non-proteinogenic amino acid \\
174.1117 $\pm$ 0.5 & + & C$_{6}$H$_{14}$N$_{4}$O$_{2}$ & arginine & biological amino acid \\
175.0845 $\pm$ 0.4 & - & C$_{7}$H$_{13}$NO$_{4}$ & spermidic acid & non-proteinogenic amino acid \\
 & & & valienamine & amino sugar \\
175.0957 $\pm$ 0.5 & - & C$_{6}$H$_{13}$N$_{3}$O$_{3}$ & citrulline & non-proteinogenic amino acid \\
176.0909 $\pm$ 8.5 & + & C$_{5}$H$_{12}$N$_{4}$O$_{3}$ & canavanine & non-proteinogenic amino acid \\
180.0634 $\pm$ 7.7 & - & C$_{6}$H$_{12}$O$_{6}$ & glucose & monosaccharide \\
 & & & hexoses & monosaccharides \\
182.079 $\pm$ 6.3 & +/- & C$_{6}$H$_{14}$O$_{6}$ & sorbitol & sugar alcohol \\
 & & & hexitols & sugar alcohols \\
185.0324 $\pm$ 6.4 & - & C$_{7}$H$_{7}$NO$_{5}$ & 2-amino-3-carboxymuconic semialdehyde & amino acid metabolite \\
188.1161 $\pm$ 0.4 & +/- & C$_{8}$H$_{16}$N$_{2}$O$_{3}$ & acetyl lysine & amino acid derivative \\
 & & & glycyl leucine & dipeptide \\
 & & & alanyl valine & dipeptide \\
188.1273 $\pm$ 0.4 & + & C$_{7}$H$_{16}$N$_{4}$O$_{2}$ & homoarginine & non-proteinogenic amino acid \\
 & & & methylarginine & amino acid derivative \\
192.111 $\pm$ 7.1 & +/- & C$_{7}$H$_{16}$N$_{2}$O$_{4}$ & valine glycine & dipeptide \\
195.0532 $\pm$ 1.1 & - & C$_{9}$H$_{9}$NO$_{4}$ & dopaquinone & amino acid derivative \\
 & & & salicyluric acid & amino acid metabolite \\
196.0583 $\pm$ 4.0 & +/- & C$_{6}$H$_{12}$O$_{7}$ & gluconic acid & sugar acid \\
 & & & hexonic acids & sugar acids \\
197.0688 $\pm$ 6.7 & - & C$_{9}$H$_{11}$NO$_{4}$ & dihydroxyphenylalanine & amino acid derivative \\
209.0688 $\pm$ 6.4 & - & C$_{10}$H$_{11}$NO$_{4}$ & hydroxyphenylacetylglycine & amino acid metabolite \\
225.0637 $\pm$ 3.2 & - & C$_{10}$H$_{11}$NO$_{5}$ & vanilloylglycine & amino acid metabolite \\
236.0797 $\pm$ 1.2 & - & C$_{11}$H$_{12}$N$_{2}$O$_{4}$ & formylkynurenine & amino acid derivative \\
246.1328 $\pm$ 1.6 & - & C$_{9}$H$_{18}$N$_{4}$O$_{4}$ & octopine & amino acid metabolite \\
270.0964 $\pm$ 1.9 & - & C$_{10}$H$_{14}$N$_{4}$O$_{5}$ & aspartyl histidine & dipeptide \\
449.3141 $\pm$ 6.8 & + & C$_{26}$H$_{43}$NO$_{5}$ & glycodeoxycholic acid & bile acid \\
465.309 $\pm$ 6.5 & + & C$_{26}$H$_{43}$NO$_{6}$ & glycocholic acid & bile acid \\
\hline
\multicolumn{1}{l}{\textit{\textbf{Destroyed}}} \\
\hline
%76.0161 $\pm$ 8.4 & - & C$_{2}$H$_{4}$O$_{3}$ &  &  \\
90.0317 $\pm$ 4.5 & - & C$_{3}$H$_{6}$O$_{3}$ & glyceraldehyde & monosaccharide \\
 & & & lactic acid & sugar derivative \\
135.0545 $\pm$ 5.8 & - & C$_{5}$H$_{5}$N$_{5}$ & adenine & nucleotide base \\
153.0426 $\pm$ 2.6 & - & C$_{7}$H$_{7}$NO$_{3}$ & aminosalicylic acids & amino acid derivative \\
 & & & hydroxyanthranilic acid & amino acid metabolite \\
\label{tab:300K_insol}
\end{longtable*}

\begin{longtable*}{c c c l l}
\caption{Molecular formulas created and destroyed after hydrolysis in the soluble fraction of the 400 K water-rich haze.} \\
\hline \hline
Mass (m/z) $\pm$ $\Delta$ppm & Mode & Formula & Potential Molecule & Relevance \\
\hline
\endfirsthead

\multicolumn{5}{c}%
{{\bfseries \tablename\ \thetable{} -- continued from previous page}} \\
\hline
Mass (m/z) $\pm$ $\Delta$ppm & Mode & Formula & Potential Molecule & Relevance \\
\hline
\endhead

\hline \multicolumn{5}{r}{{Continued on next page}} \\
\endfoot

\hline
\endlastfoot

\multicolumn{1}{l}{\textit{\textbf{Created}}} \\
\hline
102.0429 $\pm$ 3.5 & + & C$_{3}$H$_{6}$N$_{2}$O$_{2}$ & formimidoylglycine & amino acid derivative \\
 & & & cycloserine & non-proteinogenic amino acid \\
103.0633 $\pm$ 7.4 & - & C$_{4}$H$_{9}$NO$_{2}$ & aminobutyric acid & non-proteinogenic amino acid \\
 & & & dimethylglycine & amino acid derivative \\
117.079 $\pm$ 3.1 & - & C$_{5}$H$_{11}$NO$_{2}$ & valine & biological amino acid \\
 & & & trimethylglycine & amino acid derivative \\
120.0423 $\pm$ 6.7 & - & C$_{4}$H$_{8}$O$_{4}$ & erythrose & monosaccharide \\
 & & & tetroses & monosaccharides \\
129.079 $\pm$ 3.2 & - & C$_{6}$H$_{11}$NO$_{2}$ & cycloleucine & non-proteinogenic amino acid \\
132.0535 $\pm$ 4.8 & - & C$_{4}$H$_{8}$N$_{2}$O$_{3}$ & 3-ureidopropionic acid & amino acid derivative \\
 & & & glycylglycine & dipeptide \\
 & & & asparagine & biological amino acid \\
133.0375 $\pm$ 4.0 & - & C$_{4}$H$_{7}$NO$_{4}$ & iminodiacetic acid & amino acid derivative \\
 & & & aspartic acid & biological amino acid \\
133.0739 $\pm$ 0.9 & - & C$_{5}$H$_{11}$NO$_{3}$ & hydroxyaminovaleric acid & non-proteinogenic amino acid \\
134.0215 $\pm$ 2.0 & - & C$_{4}$H$_{6}$O$_{5}$ & malic acid & dicarboxylic acid \\
134.0579 $\pm$ 1.9 & - & C$_{5}$H$_{10}$O$_{4}$ & dihydroxyisovalerate & amino acid metabolite \\
 & & & deoxyribose & monosaccharide \\
136.0372 $\pm$ 3.6 & - & C$_{4}$H$_{8}$O$_{5}$ & threonic acid & sugar acid \\
 & & & erythronic acid & sugar acid \\
143.0946 $\pm$ 2.0 & - & C$_{7}$H$_{13}$NO$_{2}$ & stachydrine & amino acid derivative \\
147.0532 $\pm$ 5.2 & - & C$_{5}$H$_{9}$NO$_{4}$ & o-cetylserine & non-proteinogenic amino acid \\
 & & & glutamic acid & biological amino acid \\
149.0324 $\pm$ 8.1 & + & C$_{4}$H$_{7}$NO$_{5}$ & hydroxyaspartic acid & non-proteinogenic amino acid \\
150.0528 $\pm$ 1.6 & - & C$_{5}$H$_{10}$O$_{5}$ & ribose & monosaccharide \\
 & & & pentose sugars & monosaccharides \\
152.0685 $\pm$ 6.7 & + & C$_{5}$H$_{12}$O$_{5}$ & ribitol & sugar alcohol \\
 & & & pentitols & sugar alcohols \\
157.0375 $\pm$ 1.8 & - & C$_{6}$H$_{7}$NO$_{4}$ & 2-aminomuconic acid & non-proteinogenic amino acid \\
157.1103 $\pm$ 2.8 & - & C$_{8}$H$_{15}$NO$_{2}$ & cyclohexylglycine & non-proteinogenic amino acid \\
158.0328 $\pm$ 2.3 & - & C$_{5}$H$_{6}$N$_{2}$O$_{4}$ & dihydroorotic acid & nucleobase precursor \\
161.0688 $\pm$ 4.2 & - & C$_{6}$H$_{11}$NO$_{4}$ & $\alpha$-aminoadipic acid & non-proteinogenic amino acid \\
 & & & methylglutamate & amino acid metabolite \\
164.0685 $\pm$ 2.2 & - & C$_{6}$H$_{12}$O$_{5}$ & fucose & sugar derivative \\
 & & & deoxy hexoses & sugar derivatives \\
166.0477 $\pm$ 9.4 & - & C$_{5}$H$_{10}$O$_{6}$ & ribonic acid & sugar acid \\
 & & & pentose acids & sugar acids \\
171.1259 $\pm$ 1.2 & - & C$_{9}$H$_{17}$NO$_{2}$ & cyclohexylalanine & non-proteinogenic amino acid \\
173.0437 $\pm$ 1.6 & - & C$_{5}$H$_{7}$N$_{3}$O$_{4}$ & azaserine & non-proteinogenic amino acid \\
174.1004 $\pm$ 0.6 & +/- & C$_{7}$H$_{14}$N$_{2}$O$_{3}$ & theanine & non-proteinogenic amino acid \\
 & & & formyllysine & non-proteinogenic amino acid \\
174.1117 $\pm$ 0.3 & + & C$_{6}$H$_{14}$N$_{4}$O$_{2}$ & arginine & biological amino acid \\
175.0845 $\pm$ 1.7 & - & C$_{7}$H$_{13}$NO$_{4}$ & spermidic acid & non-proteinogenic amino acid \\
 & & & valienamine & amino sugar \\
175.0957 $\pm$ 0.5 & - & C$_{6}$H$_{13}$N$_{3}$O$_{3}$ & citrulline & non-proteinogenic amino acid \\
176.0909 $\pm$ 7.9 & + & C$_{5}$H$_{12}$N$_{4}$O$_{3}$ & canavanine & non-proteinogenic amino acid \\
180.0634 $\pm$ 0.1 & - & C$_{6}$H$_{12}$O$_{6}$ & glucose & monosaccharide \\
 & & & hexoses & monosaccharides \\
182.079 $\pm$ 1.2 & - & C$_{6}$H$_{14}$O$_{6}$ & sorbitol & sugar alcohol \\
 & & & hexitols & sugar alcohols \\
188.1161 $\pm$ 0.2 & + & C$_{8}$H$_{16}$N$_{2}$O$_{3}$ & acetyl lysine & amino acid derivative \\
 & & & glycyl leucine & dipeptide \\
 & & & alanyl valine & dipeptide \\
188.1273 $\pm$ 0.5 & + & C$_{7}$H$_{16}$N$_{4}$O$_{2}$ & homoarginine & non-proteinogenic amino acid \\
 & & & methylarginine & amino acid derivative \\
190.1066 $\pm$ 8.8 & + & C$_{6}$H$_{14}$N$_{4}$O$_{3}$ & hydroxyarginine & non-proteinogenic amino acid \\
195.0532 $\pm$ 1.1 & - & C$_{9}$H$_{9}$NO$_{4}$ & dopaquinone & amino acid derivative \\
 & & & salicyluric acid & amino acid metabolite \\
196.0484 $\pm$ 0.8 & - & C$_{8}$H$_{8}$N$_{2}$O$_{4}$ & amino nitrophenyl acetic acid & amino acid derivative \\
196.0583 $\pm$ 1.1 & +/- & C$_{6}$H$_{12}$O$_{7}$ & gluconic acid & sugar acid \\
 & & & hexonic acids & sugar acids \\
197.0688 $\pm$ 0.0 & - & C$_{9}$H$_{11}$NO$_{4}$ & dihydroxyphenylalanine & amino acid derivative \\
199.0845 $\pm$ 1.5 & - & C$_{9}$H$_{13}$NO$_{4}$ & anticapsin & non-proteinogenic amino acid \\
209.0688 $\pm$ 0.4 & - & C$_{10}$H$_{11}$NO$_{4}$ & hydroxyphenylacetylglycine & amino acid metabolite \\
%210.0641 $\pm$ 0.4 & - & C$_{9}$H$_{10}$N$_{2}$O$_{4}$ &  &  \\
211.0845 $\pm$ 0.4 & - & C$_{10}$H$_{13}$NO$_{4}$ & methoxytyrosine & non-proteinogenic amino acid \\
 & & & farinomalein & amino acid metabolite \\
219.0743 $\pm$ 6.0 & - & C$_{8}$H$_{13}$NO$_{6}$ & O-succinyl-L-homoserine & amino acid metabolite \\
%224.0797 $\pm$ 1.2 & - & C$_{10}$H$_{12}$N$_{2}$O$_{4}$ &  &  \\
225.0637 $\pm$ 0.5 & - & C$_{10}$H$_{11}$NO$_{5}$ & vanilloylglycine & amino acid metabolite \\
%236.0797 $\pm$ 1.2 & - & C$_{11}$H$_{12}$N$_{2}$O$_{4}$ &  &  \\
%246.1004 $\pm$ 1.2 & +/- & C$_{13}$H$_{14}$N$_{2}$O$_{3}$ &  &  \\
%246.1328 $\pm$ 7.0 & + & C$_{9}$H$_{18}$N$_{4}$O$_{4}$ &  &  \\
270.0964 $\pm$ 0.3 & - & C$_{10}$H$_{14}$N$_{4}$O$_{5}$ & aspartyl histidine & dipeptide \\
311.1369 $\pm$ 2.3 & - & C$_{15}$H$_{21}$NO$_{6}$ & domoic acid & non-proteinogenic amino acid \\
%342.1162 $\pm$ 3.4 & - & C$_{12}$H$_{22}$O$_{11}$ &  &  \\
\hline
\multicolumn{1}{l}{\textit{\textbf{Destroyed}}} \\
\hline
%76.0161 $\pm$ 10.0 & - & C$_{2}$H$_{4}$O$_{3}$ &  &  \\
%89.0113 $\pm$ 8.7 & - & C$_{2}$H$_{3}$NO$_{3}$ &  &  \\
%112.0273 $\pm$ 8.5 & - & C$_{4}$H$_{4}$N$_{2}$O$_{2}$ &  &  \\
%126.0429 $\pm$ 8.1 & - & C$_{5}$H$_{6}$N$_{2}$O$_{2}$ &  &  \\
%129.0426 $\pm$ 7.3 & - & C$_{5}$H$_{7}$NO$_{3}$ &  &  \\
%137.0477 $\pm$ 7.1 & - & C$_{7}$H$_{7}$NO$_{2}$ &  &  \\
%141.0426 $\pm$ 7.2 & - & C$_{6}$H$_{7}$NO$_{3}$ &  &  \\
176.0586 $\pm$ 6.2 & - & C$_{9}$H$_{8}$N$_{2}$O$_{2}$ & cyanophenyl glycine & amino acid derivative \\
\label{tab:400K_sol}
\end{longtable*}

\begin{longtable*}{c c c l l}
\caption{Molecular formulas created and destroyed after hydrolysis in the insoluble fraction of the 400 K water-rich haze.} \\
\hline \hline
Mass (m/z) $\pm$ $\Delta$ppm & Mode & Formula & Potential Molecule & Relevance \\
\hline
\endfirsthead

\multicolumn{5}{c}%
{{\bfseries \tablename\ \thetable{} -- continued from previous page}} \\
\hline
Mass (m/z) $\pm$ $\Delta$ppm & Mode & Formula & Potential Molecule & Relevance \\
\hline
\endhead

\hline \multicolumn{5}{r}{{Continued on next page}} \\
\endfoot

\hline
\endlastfoot

\multicolumn{1}{l}{\textit{\textbf{Created}}} \\
\hline
75.032 $\pm$ 6.6 & + & C$_{2}$H$_{5}$NO$_{2}$ & glycine & biological amino acid \\
89.0477 $\pm$ 4.1 & + & C$_{3}$H$_{7}$NO$_{2}$ & alanine & biological amino acid \\
 & & & sarcosine & non-proteinogenic amino acid \\
101.0477 $\pm$ 3.8 & + & C$_{4}$H$_{7}$NO$_{2}$ & azetidine-2-carboxylic acid & non-proteinogenic amino acid \\
 & & & vinyl glycine & non-proteinogenic amino acid \\
102.0429 $\pm$ 3.5 & + & C$_{3}$H$_{6}$N$_{2}$O$_{2}$ & formimidoylglycine & amino acid derivative \\
 & & & cycloserine & non-proteinogenic amino acid \\
103.0633 $\pm$ 3.8 & - & C$_{4}$H$_{9}$NO$_{2}$ & aminobutyric acid & non-proteinogenic amino acid \\
 & & & dimethylglycine & amino acid derivative \\
104.0474 $\pm$ 2.7 & + & C$_{4}$H$_{8}$O$_{3}$ & hydroxybutyric acid & hydroxy acid \\
115.0633 $\pm$ 2.9 & - & C$_{5}$H$_{9}$NO$_{2}$ & proline & biological amino acid \\
 & & & allylglycine & amino acid derivative \\
117.079 $\pm$ 2.4 & - & C$_{5}$H$_{11}$NO$_{2}$ & valine & biological amino acid \\
 & & & trimethylglycine & amino acid derivative \\
118.0266 $\pm$ 3.2 & + & C$_{4}$H$_{6}$O$_{4}$ & succinic acid & dicarboxylic acid \\
118.0742 $\pm$ 2.5 & + & C$_{4}$H$_{10}$N$_{2}$O$_{2}$ & diaminobutyric acid & non-proteinogenic amino acid \\
 & & & $\beta$-methylamino-L-alanine & non-proteinogenic amino acid \\
119.0582 $\pm$ 3.3 & + & C$_{4}$H$_{9}$NO$_{3}$ & $\gamma$-amino-$\beta$-hydroxybutyric acid & hydroxy acid derivative \\
 & & & homoserine & non-proteinogenic amino acid \\
 & & & threonine & biological amino acid \\
120.0423 $\pm$ 1.2 & - & C$_{4}$H$_{8}$O$_{4}$ & erythrose & monosaccharide \\
 & & & tetroses & monosaccharides \\
122.0579 $\pm$ 8.1 & - & C$_{4}$H$_{10}$O$_{4}$ & erythritol & sugar alcohol \\
 & & & threitol & sugar alcohol \\
129.079 $\pm$ 1.6 & + & C$_{6}$H$_{11}$NO$_{2}$ & cycloleucine & non-proteinogenic amino acid \\
132.0535 $\pm$ 0.9 & - & C$_{4}$H$_{8}$N$_{2}$O$_{3}$ & 3-ureidopropionic acid & amino acid derivative \\
 & & & glycylglycine & dipeptide \\
 & & & asparagine & biological amino acid \\
133.0375 $\pm$ 1.1 & - & C$_{4}$H$_{7}$NO$_{4}$ & iminodiacetic acid & amino acid derivative \\
 & & & aspartic acid & biological amino acid \\
134.0579 $\pm$ 2.2 & - & C$_{5}$H$_{10}$O$_{4}$ & dihydroxyisovalerate & amino acid metabolite \\
 & & & deoxyribose & monosaccharide \\
146.0691 $\pm$ 0.4 & - & C$_{5}$H$_{10}$N$_{2}$O$_{3}$ & $\beta$-ureidoisobutyric acid & amino acid metabolite \\
 & & & glutamine & biological amino acid \\
147.0532 $\pm$ 1.9 & - & C$_{5}$H$_{9}$NO$_{4}$ & o-cetylserine & non-proteinogenic amino acid \\
 & & & glutamic acid & biological amino acid \\
150.0528 $\pm$ 3.0 & +/- & C$_{5}$H$_{10}$O$_{5}$ & ribose & monosaccharide \\
 & & & pentose sugars & monosaccharides \\
152.0685 $\pm$ 6.7 & + & C$_{5}$H$_{12}$O$_{5}$ & ribitol & sugar alcohol \\
 & & & pentitols & sugar alcohols \\
157.0375 $\pm$ 0.6 & + & C$_{6}$H$_{7}$NO$_{4}$ & 2-aminomuconic acid & non-proteinogenic amino acid \\
157.1103 $\pm$ 3.9 & + & C$_{8}$H$_{15}$NO$_{2}$ & cyclohexylglycine & non-proteinogenic amino acid \\
158.0328 $\pm$ 0.3 & +/- & C$_{5}$H$_{6}$N$_{2}$O$_{4}$ & dihydroorotic acid & nucleobase precursor \\
159.1259 $\pm$ 4.3 & + & C$_{8}$H$_{17}$NO$_{2}$ & aminooctonoic acid & amino acid metabolite \\
 & & & valerobetaine & amino acid metabolite \\
161.0688 $\pm$ 0.9 & - & C$_{6}$H$_{11}$NO$_{4}$ & $\alpha$-aminoadipic acid & non-proteinogenic amino acid \\
 & & & methylglutamate & amino acid metabolite \\
164.0685 $\pm$ 3.7 & +/- & C$_{6}$H$_{12}$O$_{5}$ & fucose & sugar derivative \\
 & & & deoxy hexoses & sugar derivatives \\
166.0477 $\pm$ 7.9 & +/- & C$_{5}$H$_{10}$O$_{6}$ & ribonic acid & sugar acid \\
 & & & pentose acids & sugar acids \\
173.0437 $\pm$ 0.3 & +/- & C$_{5}$H$_{7}$N$_{3}$O$_{4}$ & azaserine & non-proteinogenic amino acid \\
174.1004 $\pm$ 0.6 & +/- & C$_{7}$H$_{14}$N$_{2}$O$_{3}$ &  theanine & non-proteinogenic amino acid \\
 & & & formyllysine & non-proteinogenic amino acid \\
175.0845 $\pm$ 0.3 & +/- & C$_{7}$H$_{13}$NO$_{4}$ & spermidic acid & non-proteinogenic amino acid \\
 & & & valienamine & amino sugar \\
175.0957 $\pm$ 9.0 & - & C$_{6}$H$_{13}$N$_{3}$O$_{3}$ & citrulline & non-proteinogenic amino acid \\
176.0909 $\pm$ 8.7 & - & C$_{5}$H$_{12}$N$_{4}$O$_{3}$ & canavanine & non-proteinogenic amino acid \\
180.0634 $\pm$ 6.6 & + & C$_{6}$H$_{12}$O$_{6}$ & glucose & monosaccharide \\
 & & & hexoses & monosaccharides \\
182.079 $\pm$ 5.4 & - & C$_{6}$H$_{14}$O$_{6}$ & sorbitol & sugar alcohol \\
 & & & hexitols & sugar alcohols \\
185.0324 $\pm$ 0.5 & + & C$_{7}$H$_{7}$NO$_{5}$ & 2-amino-3-carboxymuconic semialdehyde & amino acid metabolite \\
188.1161 $\pm$ 0.2 & + & C$_{8}$H$_{16}$N$_{2}$O$_{3}$ & acetyl lysine & amino acid derivative \\
 & & & glycyl leucine & dipeptide \\
 & & & alanyl valine & dipeptide \\
195.0532 $\pm$ 0.0 & +/- & C$_{9}$H$_{9}$NO$_{4}$ & dopaquinone & amino acid derivative \\
 & & & salicyluric acid & amino acid metabolite \\
196.0484 $\pm$ 0.2 & +/- & C$_{8}$H$_{8}$N$_{2}$O$_{4}$ & amino nitrophenyl acetic acid & amino acid derivative \\
196.0583 $\pm$ 7.1 & - & C$_{6}$H$_{12}$O$_{7}$ & gluconic acid & sugar acid \\
 & & & hexonic acids & sugar acids \\
197.0688 $\pm$ 0.2 & +/- & C$_{9}$H$_{11}$NO$_{4}$ & dihydroxyphenylalanine & amino acid derivative \\
199.0845 $\pm$ 0.0 & + & C$_{9}$H$_{13}$NO$_{4}$ & anticapsin & non-proteinogenic amino acid \\
209.0688 $\pm$ 0.2 & - & C$_{10}$H$_{11}$NO$_{4}$ & hydroxyphenylacetylglycine & amino acid metabolite \\
%210.0641 $\pm$ 0.1 & + & C$_{9}$H$_{10}$N$_{2}$O$_{4}$ &  &  \\
211.0845 $\pm$ 0.4 & +/- & C$_{10}$H$_{13}$NO$_{4}$ & methoxytyrosine & non-proteinogenic amino acid \\
 & & & farinomalein & amino acid metabolite \\
215.0946 $\pm$ 0.6 & - & C$_{13}$H$_{13}$NO$_{2}$ & napthylalanine & non-proteinogenic amino acid \\
219.0743 $\pm$ 0.8 & +/- & C$_{8}$H$_{13}$NO$_{6}$ & O-succinyl-L-homoserine & amino acid metabolite \\
%224.0797 $\pm$ 0.0 & + & C$_{10}$H$_{12}$N$_{2}$O$_{4}$ &  &  \\
225.0637 $\pm$ 0.3 & + & C$_{10}$H$_{11}$NO$_{5}$ & vanilloylglycine & amino acid metabolite \\
%236.0797 $\pm$ 0.2 & +/- & C$_{11}$H$_{12}$N$_{2}$O$_{4}$ &  &  \\
%246.1004 $\pm$ 0.5 & - & C$_{13}$H$_{14}$N$_{2}$O$_{3}$ &  &  \\
270.0964 $\pm$ 1.7 & - & C$_{10}$H$_{14}$N$_{4}$O$_{5}$ & aspartyl histidine & dipeptide \\
311.1369 $\pm$ 2.0 & - & C$_{15}$H$_{21}$NO$_{6}$ & domoic acid & non-proteinogenic amino acid \\
\hline
\multicolumn{1}{l}{\textit{\textbf{Destroyed}}} \\
\hline
––– & & & & \\
%76.0161 $\pm$ 10.0 & - & C$_{2}$H$_{4}$O$_{3}$ &  &  \\
%90.0317 $\pm$ 8.9 & - & C$_{3}$H$_{6}$O$_{3}$ &  &  \\
\label{tab:400K_insol}
\end{longtable*}

\end{document}